\documentclass[aps,prd,reprint,amsmath,nobibnotes,showpacs]{revtex4-1}

\usepackage[dvips]{graphicx}
\printfigures
\usepackage{dcolumn}
\usepackage{amsfonts}
\usepackage{amsmath}
\usepackage{bm}
\usepackage{hyperref}
\begin{document}
\title{Modeling interacting dark energy models with Chebyshev polynomials:\\Exploring their constraints and effects}
\author{Freddy Cueva Solano}
\affiliation{Instituto de F\'{\i}sica y Matem\'aticas, Universidad Michoacana de San Nicol\'as de Hidalgo\\
Edificio C-3, Ciudad Universitaria, CP. 58040, Morelia, Michoac\'an, M\'exico.}
\email{freddy@ifm.umich.mx,\;\;freddycuevasolano$2009$@gmail.com}
\begin{abstract}
In this work, we examine the main cosmological effects derived from a time-varying coupling ($\bar{Q}$) between a dark matter (DM) fluid and a dark energy (DE) fluid with 
time-varying DE equation of state (EoS) parameter ($\mathrm{\omega_{DE}}$), in two different coupled DE models. These scenarios were built in terms of Chebyschev polynomials. 
Our results show that such models within dark sector can suffer an instability in their perturbations at early times and a slight departure on the amplitude of the cosmic
structure growth ($\mathrm{f\sigma_{8}}$) from the standard background evolution of the matter. These effects depend on the form of both $\bar{Q}$ and $\mathrm{\omega_{DE}}$. Here, we also 
perform a combined statistical analysis using current data to put tighter constraints on the parameters space. Finally, we use some selections criteria to distinguish our 
models.
\end{abstract}
\pacs{98.80.-k, 95.35.+d, 95.36.+x, 98.80.Es} 
\maketitle 
\section{Introduction} 
A number of observations \cite{Conley2011,Jonsson2010,Betoule2014,Jackson1972,Kaiser1987,Mehrabi2015,Alcock1979,Seo2008,Battye2015,Samushia2014,Hudson2013,Beutler2012,
Feix2015,Percival2004,Song2009,Tegmark2006,Guzzo2008,Samushia2012,Blake2011,Tojeiro2012,Reid2012,delaTorre2013,Planck2015,Hinshaw2013,Beutler2011,Ross2015,Percival2010,
Kazin2010,Padmanabhan2012,Chuang2013a,Chuang2013b,Anderson2014a,Kazin2014,Debulac2015,FontRibera2014,Eisenstein1998,Eisenstein2005,Hemantha2014,Bond-Tegmark1997,
Hu-Sugiyama1996,Neveu2016,Sharov2015,Zhang2014,Simon2005,Moresco2012,Gastanaga2009,Oka2014,Blake2012,Stern2010,Moresco2015,Busca2013} 
have indicated that the present universe is undergoing a phase of accelerated expansion, and driven probably by a new form of energy with negative EoS parameter, commonly 
so-called DE \cite{DES2006}. This energy has been interpreted in various forms and extensely studied in \cite{OptionsDE}. 
However, within General Relativity (GR) the DE models can suffer the coincidence problem, namely why the DM and DE energy densities are of the same order today. 
This latter problem could be solved or even alleviated, by assuming the existence of a non-gravitational $\bar{Q}$ within the dark sector, which gives rise to a continuous 
energy exchange from DE to DM or vice-versa. Currently, there are not neither physical arguments nor recent observations to exclude $\bar{Q}$ 
\cite{Interacting,Pavons,Wangs,Cueva-Nucamendi2012}. Moreover, due to the absence of a fundamental theory to construct $\bar{Q}$, different ansatzes have been widely 
discussed in \cite{Interacting,Pavons,Wangs,Cueva-Nucamendi2012}. So, It has been shown in some coupled DE scenarios that an appropriate choice of $\bar{Q}$ may cause 
serious instabilities in the dark sector perturbations at early times \cite{valiviita2008,Jackson2009,He2009,Xu2011,He2011,Clemson2012}. Also, It has been argued in the 
coupled DE scenarios that $\bar{Q}$ can affect the background evolution of the DM density perturbations and the expansion history of the universe \cite{Mehrabi2015, Alcaniz2013, Yang2014}. Thus, $\bar{Q}$ and 
$\mathrm{\omega_{DE}}$ could very possibly introduce new features on $\mathrm{f\sigma_{8}}$ during the matter era.\\
On the other one, within dark sector we can propose new ansatzes for both $\bar{Q}$ and $\mathrm{\omega_{DE}}$, which can be expanded in terms of the Chebyshev polynomials $T_{n}$, 
defined in the interval $[-1,1]$ and with a diverge-free $\mathrm{\omega_{DE}}$ at z$\rightarrow -1$ \cite{Chevallier-Linder, Li-Ma}. However, that polynomial base was particularly 
chosen due to its rapid convergence and better stability than others, by giving minimal errors \cite{Simon2005,Martinez2008}. Besides, $\bar{Q}$ could also be proportional 
to the DM energy density $\mathrm{{\bar{\rho}}_{DM}}$ and to the Hubble parameter $\mathrm{\bar{H}}$. This new model will guarantee an accelerated scaling attractor and connect to a standard 
evolution of the matter. Here, $\bar{Q}$ will be restricted from the criteria exhibit in \cite{Campo-Herrera2015}.\\
Therefore, our main motivation in the present work is to explore the effects of $\bar{Q}$ and $\mathrm{\omega_{DE}}$ on the cosmological variables in differents times and scales, 
including the search for new ways to alleviate the coincidence problem.\\
On the other hand, two distinct coupled DE models (XCPL and DR) are discussed, on which we have performed a global fitting using an analysis combined of Joint Light Curve 
Analysis (JLA) type Ia Supernovae (SNe Ia) data \cite{Conley2011,Jonsson2010,Betoule2014}, including the growth rate of structure formation obtained from 
redshift space distortion (RSD) data \cite{Jackson1972,Kaiser1987,Mehrabi2015,Alcock1979,Seo2008,Battye2015,Samushia2014,Hudson2013,Beutler2012,Feix2015,Percival2004,Song2009,
Tegmark2006,Guzzo2008,Samushia2012,Blake2011,Tojeiro2012,Reid2012,delaTorre2013,Planck2015}, together with Baryon Acoustic Oscillation (BAO) data \cite{Hinshaw2013,
Beutler2011,Ross2015,Percival2010,Kazin2010,Padmanabhan2012,Chuang2013a,Chuang2013b,Anderson2014a,Kazin2014,Debulac2015,FontRibera2014,Eisenstein1998,Eisenstein2005,
Hemantha2014}, as well as the observations of anisotropies in the power spectrum of the Cosmic Microwave Background (CMB) data \cite{Planck2015,
Bond-Tegmark1997,Hu-Sugiyama1996,Neveu2016} and the Hubble parameter (H) data obtained from galaxy surveys \cite{Sharov2015,Zhang2014,Simon2005,Moresco2012,Gastanaga2009,
Oka2014,Blake2012,Stern2010,Moresco2015,Busca2013} to constrain the parameter space of such models and break the degeneracy of their parameters, putting tighter 
constraints on them.\\
In this paper, we will use the following criteria of selection $\chi^{2}/dof$ ($dof$: degrees of freedom), Goodness of Fit ($GoF$), Akaike Information 
Criterion ($AIC$) \cite{AIC_Criterion}, Bayesian Information Criterion ($BIC$) \cite{BIC_Criterion,Kurek2014,Arevalo2016} to distinguish our cosmological models from the 
number of their parameters that require to explain the data.\\
Finally, we organize this paper as follows: We describe the background equation of motions of the universe in Sec. II and the perturbed universe in Sec. III. The cu-rrent 
observational data and the priors considered are presented in Sec. IV. In Sec. V, we describe the different selection criteria. We discuss our results in Sec. VI and show 
our conclusions in Sec. VII.
\section{Background equations of motion}\label{Background}
We consider here a flat Friedmann-Robertson-Walker (FRW) universe composed by radiation, baryons, DM and DE. Moreover, we postulate that the dark components can also 
interact through a non-gravitational coupling $\bar{Q}$. For $\bar{Q}>0$ ($\bar{Q}<0$) implies that the energy flows from DE to DM (the energy flows from DM to DE). 
Also, to satisfy the requirements imposed by local gravity experiments \cite{Koyama2009-Brax2010} we also assume that baryons 
and radiation are coupled to the dark components only through the gravity. Thus, the energy balance equations for these fluids may be described \cite{Cueva-Nucamendi2012}, 
\begin{eqnarray}
\label{EB}\frac{\mathrm{d\bar{\rho}_{b}}}{\mathrm{dz}}-3\mathrm{{\bar{H}}{\bar{\rho}}_{b}}&=&0\,,\\
\label{Er}\frac{\mathrm{d\bar{\rho}_{r}}}{\mathrm{dz}}-4\mathrm{{\bar{H}}{\bar{\rho}}_{r}}&=&0\,,\\
\label{EDM}\frac{\mathrm{d\bar{\rho}_{DM}}}{\mathrm{dz}}-\frac{3\mathrm{{\bar{\rho}}_{DM}}}{(1+\mathrm{z})}&=&-\frac{\bar{Q}}{\mathrm{\bar{H}(1+z)}}\,,\\
\label{EDE}\frac{\mathrm{d\bar{\rho}_{DE}}}{\mathrm{dz}}-\frac{3(1+\mathrm{\omega_{DE}})\mathrm{{\bar{\rho}}_{DE}}}{(1+\mathrm{z})}&=&+\frac{\bar{Q}}{\mathrm{\bar{H}(1+z)}}\,,
\end{eqnarray}
where $\mathrm{{\bar{\rho}}_{b}}$, $\mathrm{{\bar{\rho}}_{r}}$, $\mathrm{{\bar{\rho}}_{DM}}$ and $\mathrm{{\bar{\rho}}_{DE}}$ are the energy densities of the baryon (b), radiation (r), DM and DE, 
respectively, and $\mathrm{\omega_{DE}}=P_{DE}/\mathrm{\bar{\rho}_{DE}}<0$.\\
We also defined the critical density $\rho_{c}\equiv 3{\mathrm{\bar{H}}}^2/8\pi G$ and the critical density today $\rho_{c,0}\equiv 3{\mathrm{H}}^2_{0}/8\pi G$ (here $\mathrm{H_{0}}$ is the current value of the Hubble parameter). Considering that $A=b, r, DM, DE$, the normalized densities are
\begin{equation}\label{energydensity} 
\mathrm{{\bar{\Omega}}_{A}}\equiv\frac{\mathrm{\bar{\rho}_{A}}}{\rho_{c}}=\frac{\mathrm{\bar{\rho}_{A}}/\rho_{c,0}}{\rho_{c}/\rho_{c,0}}=\frac{\bar{\Omega}^{\star}_{A}}{E^{2}}\,,\qquad{\Omega}_{A,0}\equiv
\frac{{\rho}_{A,0}}{\rho_{c,0}}\;,
\end{equation}
and the first Friedmann equation is given by
\begin{eqnarray}\label{hubble} 
E^{2}&\equiv&\frac{\mathrm{\bar{H}}^{2}}{\mathrm{{H}^{2}_{0}}}=\frac{8\pi G}{3{\mathrm{H}}^{2}_{0}}\left(\mathrm{\bar{\rho}_{b}}+\mathrm{{\bar{\rho}}_{r}}+\mathrm{{\bar{\rho}}_{DM}}+\mathrm{{\bar{\rho}}_{DE}}\right)\;,\nonumber\\
&& =\left[{\Omega}^{\star}_{b}+{\Omega}^{\star}_{r}+{\Omega}^{\star}_{DM}+{\Omega}^{\star}_{DE}\right]\,.
\end{eqnarray}
The following relation is valid for all time
\begin{equation}\label{OmegaDE_present} 
\mathrm{{\bar{\Omega}}_{b}}+\mathrm{{\bar{\Omega}}_{r}}+\mathrm{{\bar{\Omega}}_{DM}}+\mathrm{{\bar{\Omega}}_{DE}}=1\,\,.  
\end{equation}
\subsection{Parameterizations of $\bar{Q}$ and $\mathrm{\omega_{DE}}$} \label{General Reconstruction}
Due to the fact that the origin and nature of the dark fluids are unknown, it is not possible to derive $\bar{Q}$ from fundamental principles. However, we have the freedom of 
choosing any possible form of $\bar{Q}$ that satisfies Eqs. (\ref{EDM}) and (\ref{EDE}) simultaneously. Hence, we propose a new phenomenological form for a varying $\bar{Q}$ 
so that it can alleviate the coincidence problem. This coupling could be chosen proportional to $\mathrm{{\bar{\rho}}_{DM}}$, to $\mathrm{\bar{H}}$ and to ${\bar{\rm I}}_{\rm Q}$. 
Therefore, $\bar{Q}$ in the dark sector can be written as
\begin{equation}\label{Interaction}
{\bar{Q}}\equiv \mathrm{\bar{H}}\mathrm{\bar{\rho}_{DM}}\bar{{\rm I}}_{\rm Q}\;,\qquad {\bar{\rm I}}_{\rm Q}\equiv \sum_{n=0}^{2}{{\lambda}}_{n}T_{n}\;.
\end{equation}
Here ${\bar{\rm I}}_{\rm Q}$ measures the strength of the coupling, and was modeled as a varying function of $\mathrm{z}$ in terms of Chebyshev polynomials.
This polynomial base was chosen because it converges rapidly, is more stable than others and behaves well in any polynomial expansion, giving minimal errors. 
The coefficients ${\lambda}_{n}$ are free dimensionless parameters \cite{Cueva-Nucamendi2012} and the first three Chebyshev polynomials are
\begin{equation}\label{Chebyshev1} 
T_{0}(z)=1\;,\hspace{0.3cm}T_{1}(z)=z\;,\hspace{0.3cm} T_{2}(z)=(2z^{2}-1)\;. 
\end{equation}
Similarly, we propose here a new phenomenological ansatz for a time-varying $\mathrm{\omega_{DE}}$ and divergence-free at $\mathrm{z}\rightarrow -1$. Thus, we can write  
\begin{equation}\label{CouplingIw} 
\mathrm{\omega_{DE}}(\mathrm{z})\equiv \omega_{2}+2\sum^{2}_{m=0}\frac{\omega_{m}T_{m}}{2+{\mathrm{z}}^{2}}\,,
\end{equation}
where $\omega_{0}, \omega_{1}$ and $\omega_{2}$ are free dimensionless parameters. The polynomial $(2+{\mathrm{z}}^{2})^{-1}$ and the parameter $\omega_{2}$ 
were included conveniently to simplify the calculations. Here, $\mathrm{{\omega}_{DE}}$ behaves nearly linear at low redshift $\mathrm{{\omega}_{DE}(z=0)}=\omega_{0}$ and 
$d{\omega}/dz|_{z=0}=\omega_{1}$, while for $z\gg 1$, $\mathrm{{\omega}_{DE}(z)} \simeq 5{\omega_{2}}$.\\
\subsection{Ratios between the DM and DE energy densities}\label{Ratios}
From Eqs. (\ref{EDM}) and (\ref{EDE}), we define $\mathrm{R}\equiv \mathrm{{\bar{\rho}}_{DM}}/\mathrm{{\bar{\rho}}_{DE}}$ as the ratio of the 
energy densities of DM and DE. Then, we can rewrite $\bar{Q}$ as \cite{Campo-Herrera2015,Ratios}
\begin{equation}\label{FormQ}
\bar{Q}=-\left(3\mathrm{\omega_{DE}}+\frac{\mathrm{dR}}{\mathrm{dz}}\frac{(1+\mathrm{z})}{\mathrm{R}}\right)\frac{\mathrm{H{\bar{\rho}}_{DM}}}{1+\mathrm{R}}\,\,.
\end{equation}
This leads to the evolution equation of $\mathrm{R}$
\begin{equation}\label{ratio}
\frac{\mathrm{dR}}{\mathrm{dz}}=\frac{-\mathrm{R}}{(1+\mathrm{z})}\left(3\mathrm{\omega_{DE}}+\frac{(1+\mathrm{R})\bar{Q}}{\mathrm{\bar{H}\rho_{DM}}}\right)\;.
\end{equation}
Now, we substitute Eq. (\ref{Interaction}) into Eq. (\ref{ratio}), and then, we impose the following condition ${\mathrm{d}R}/{\mathrm{d}z}=0$ 
to guarantee the possibility that the coupling can solve the coincidence problem. This implies two solutions $\mathrm{R_{+}}=\mathrm{R(z\rightarrow\infty)}$ and $\mathrm{R_{-}}=\mathrm{R(z\rightarrow -1)}$,
\begin{equation}
\label{Stationary1}
\mathrm{R_{+}}=-\left(1+\frac{3\mathrm{\omega_{DE}}}{{\bar{\rm I}}_{\rm Q}}\right)\,,\qquad \mathrm{R_{-}}=0\,.
\end{equation}
From Eq. (\ref{Stationary1}), we note that $\mathrm{R_{+}}$ and ${{\bar{\rm I}}_{\rm Q}}$ are not independent but its product can be approached to the order unity. So, 
\begin{equation}\label{approach}
{{\bar{\rm I}}_{\rm Q}}\mathrm{R_{+}} \sim -3\mathrm{\omega_{DE}}. 
\end{equation}
In the limiting cases $\mathrm{z}\rightarrow\infty$, $\mathrm{z}\rightarrow 0$ and $\mathrm{z}\rightarrow -1$, $\mathrm{R}$ must be either constant or very slowly. Then, the quantities $\mathrm{R_{+}}$, $\mathrm{R_{-}}$ and 
the value of $\mathrm{R}$ today ($\mathrm{R_{0}}$) must fulfill $0\leq \mathrm{R_{-}}<\mathrm{R_{0}}<1\ll|\mathrm{R_{+}}|$. 
\begin{widetext}
\subsection{DE\,models}\label{DE_models} 
\subsubsection{$\Lambda$CDM model}\label{LCDM}
Fixing both $\mathrm{\omega_{DE}(z)}=-1$ and $\bar{Q}(z)=0$ into Eqs. (\ref{EB})-(\ref{EDE}) and solving Eq. (\ref{ratio}), we can find $E^{2}$ and $\mathrm{R}$ 
\begin{align}\label{hubble_LCDM}
E^{2}(\mathrm{z})&=\biggl[\Omega_{b,0}{(1+\mathrm{z})}^{3}+\Omega_{r,0}{(1+\mathrm{z})}^{4}+\Omega^{\star}_{DM}(\mathrm{z})+\Omega_{DE,0}\biggr],& \mathrm{R(z)}&=\mathrm{R_{0}}(1+\mathrm{z})^{3},& \Omega^{\star}_{DM}(\mathrm{z})&=\Omega_{DM,0}{(1+\mathrm{z})}^{3}.
\end{align}
\subsubsection {CPL model}\label{CPL} 
Replacing both $\mathrm{\omega_{DE}(z)}=\omega_{0}+\omega_{1}\mathrm{[z/(1+z)]}$, where $\omega_{0}$, $\omega_{1}$ are real parameters and $\bar{Q}(\mathrm{z})=0$ into Eqs. (\ref{EB})-(\ref{EDE}), 
and solving Eq. (\ref{ratio}), we obtain $E^{2}$ and $\mathrm{R}$, 
\begin{align}\label{hubble_CPL}
E^{2}(\mathrm{z})&=\biggl[\Omega_{b,0}{(1+\mathrm{z})}^{3}+\Omega_{r,0}{(1+\mathrm{z})}^{4}+\Omega^{\star}_{DM}(\mathrm{z})(1+\frac{1}{\mathrm{R}})\biggr],& \mathrm{R(z)}&=\mathrm{R_{0}}\left(1+\mathrm{z}\right)^{-3(\omega_{0}+\omega_{1})}{\rm exp}{\biggl(\frac{3\omega_{1}\mathrm{z}}{1+\mathrm{z}}\biggr)},
\end{align}
where $\Omega^{\star}_{DM}$ is given by Eq. (\ref{hubble_LCDM}).
\subsubsection {XCPL model}\label{XCPL} 
Putting $\mathrm{\omega_{DE}(z)}=\omega_{0}+\omega_{1}\mathrm{(z/(1+z))}$, where $\omega_{0}$, $\omega_{1}$ are real free parameters and using Eq. 
(\ref{Interaction}) into Eqs. (\ref{EB})-(\ref{EDE}), then the solution of Eq. (\ref{ratio}) gives 
\begin{align}\label{radio_XCPL}
E^{2}(\mathrm{z})&=\biggl[\Omega_{b,0}{(1+\mathrm{z})}^{3}+\Omega_{r,0}{(1+\mathrm{z})}^{4}+\Omega^{\star}_{DM}(\mathrm{z})(1+\frac{1}{\mathrm{R}})\biggr]\,,\quad \Omega^{\star}_{DM}(\mathrm{z})=(1+\mathrm{z})^{3}{\Omega_{DM,0}}{\rm exp}\biggl[{\frac{-z_{max}}{2}\sum_{n=0}^{2}\lambda_{n}I_{n}(\mathrm{z})}\biggr]\,,\nonumber\\
\mathrm{R}&=\left(\mathrm{R_{0}}R_{1}+R_{2}\right)^{-1}\,,\quad R_{1}={\rm exp}{\biggl[{\frac{z_{max}}{2}\sum_{n=0}^{2}\lambda_{n}I_{n}(\mathrm{z})}-
\frac{3\omega_{1}\mathrm{z}}{1+\mathrm{z}}\biggr]}(1+\mathrm{z})^{3(\omega_{o}+\omega_{1})}\,,\quad R_{2}=R_{21}R_{22}R_{23}\,,
\nonumber\\
R_{21}&={|(1+\mathrm{z})|}^{(\lambda_{0}-\lambda_{1}+\lambda_{2}+3(\omega_{o}+\omega_{1}))}\,,\quad 
R_{22}={\rm exp}{\biggl[-(4\lambda_{2}-\lambda_{1})\mathrm{z}+\frac{3\omega_{1}}{1+\mathrm{z}}+\lambda_{2}(1+\mathrm{z})^{2}\biggr]}\,,\nonumber\\ 
R_{2a}&=\left(\lambda_{0}+\lambda_{1}\tilde{x}+\lambda_{2}(2\tilde{x}^{2}-1)\right)\;,\quad 
R_{2b}={{|1+0.5z_{max}(1+\tilde{x})|}^{-(\lambda_{0}-\lambda_{1}+\lambda_{2}+3(\omega_{o}+\omega_{1})+1)}}\,,\nonumber\\ 
R_{2c}&={\rm exp}{\biggl[0.5z_{max}(1+\tilde{x})(4\lambda_{2}-\lambda_{1})-{3\omega_{1}}{({1+0.5z_{max}(1+\tilde{x})})}^{-1}\biggr]}\,,
\nonumber\\ 
R_{2d}&={\rm exp}{\biggl[-\lambda_{2}\left(1+z_{max}(1+\tilde{x})+0.25z^{2}_{max}(1+\tilde{x})^{2}\right)\biggr]}\;,\quad 
R_{23}=\frac{z_{max}}{2}\int_{-1}^{2(\mathrm{z}/z_{max})-1}\left(R_{2a}R_{2b}R_{2c}R_{2d}\right)d\tilde{x}\,,
\nonumber\\
\int_{0}^{\mathrm{z}}\frac{T_{n}(\tilde{x})}{(1+\tilde{x})}d\tilde{x}&\approx\frac{z_{max}}{2}\int_{-1}^{x}\frac{T_{n}(\tilde{x})}{(a_{1}+a_{2}\tilde{x})}d\tilde{x}\equiv\frac{z_{max}}{2}I_{n}(\mathrm{z})\,,\quad x\equiv 2(\frac{\mathrm{z}}{z_{max}})-1\,,\quad a_{1}\equiv1+\frac{z_{max}}{2}\,,\quad a_{2}\equiv\frac{z_{max}}{2}\,,
\nonumber\\
I_{0}(\mathrm{z})&=\frac{2}{z_{max}}\ln(1+\mathrm{z})\;,\quad I_{1}(\mathrm{z})=\frac{2}{z_{max}}\biggl[\frac{2\mathrm{z}}{z_{max}}-\frac{(2+z_{max})}{z_{max}}\ln(1+\mathrm{z})
\biggr]\;,
\nonumber\\
I_{2}(\mathrm{z})&=\frac{2}{z_{max}}\biggl[\frac{4 \mathrm{z}}{z_{max}}\left(\frac{\mathrm{z}}{z_{max}}-\frac{2}{z_{max}}-2\right)+ \left(1+\frac{6.8284}{z_{max}}\right)\left(1+\frac{1.1716}{z_{max}}\right)\ln(1+\mathrm{z})\biggr]\,,
\end{align}
where $z_{max}$ is the maximum value of $\mathrm{z}$ such that $\tilde{x}\in [-1, 1]$ and $\vert{T_{n}}(\tilde{x})\vert\leq1$ and $n\in[0,2]$ (see \cite{Cueva-Nucamendi2012}).\\ 
\subsubsection{DR model}\label{DR}
This scenario can be modeled setting Eqs. (\ref{Interaction}) and (\ref{CouplingIw}) into Eqs. (\ref{EB})-(\ref{EDE}), and then solving Eq. (\ref{ratio}), we get 
\begin{align}\label{radio_DR} 
E^{2}(\mathrm{z})&=\biggl[\Omega_{b,0}{(1+\mathrm{z})}^{3}+\Omega_{r,0}{(1+\mathrm{z})}^{4}+\Omega^{\star}_{DM}(\mathrm{z})(1+\frac{1}{\mathrm{R}})\biggr]\,,\quad \Omega^{\star}_{DM}(\mathrm{z})=(1+\mathrm{z})^{3}{\Omega_{DM,0}}{\rm exp}\biggl[{\frac{-z_{max}}{2}\sum_{n=0}^{2}\lambda_{n}I_{n}(\mathrm{z})}\biggr]\,,\nonumber\\
R&=\left(R_{0}R_{1}+R_{2}\right)^{-1}\,,\quad R_{1}=C_{0}C_{1}C_{2}\,,\quad R_{2}=R_{21}R_{22}\,,\quad R_{21}=D_{0}D_{1}D_{2}\;,\quad R_{22}=\frac{z_{max}}{2}\int_{-1}^{2(\mathrm{z}/z_{max})-1}\left(F_{0}F_{1}F_{2}\right)d\tilde{x}\,.\nonumber\\
C_0(\mathrm{z})&=(1+\mathrm{z})^{(3\omega_{2}+A_0)}{\rm exp}\biggl[\sum_{n=0}^{2}\left(\frac{z_{max}\lambda_{n}I_{n}(\mathrm{z})}{2}\right)\biggr]\,,\quad C_1(\mathrm{z})=\biggl[\frac{(2\mathrm{z}-z_{max})^{2}+2(z_{max}+2)^{2}}{3z_{max}^{2}+8(1+z_{max})}\biggr]^{A_1}\;,\nonumber\\
C_2(\mathrm{z})&={\rm exp}\Biggl[A_{2}\sqrt{2}\left(\arctan\left[\frac{\sqrt{2}[2\mathrm{z}-z_{max}]}{2z_{max}+4}\right]+\arctan\left[\frac{\sqrt2z_{max}}{2z_{max}+4}\right]\right)\Biggr]\;,\nonumber\\
D_0(\mathrm{z})&=0.5z_{max}(1+\mathrm{z})^{({\lambda_{0}-\lambda_{1}+\lambda_{2}+2(\omega_{0}-\omega_{1})+5\omega_{2}})}(2+\mathrm{z}^{2})^{(\omega_{1}-\omega_{0}+5\omega_{2})}\;,\nonumber\\
D_1(\mathrm{z})&={\rm exp}\biggl[{\sqrt{2}}\biggl(\omega_{0}+2\omega_{1}-5\omega_{2}\biggr)\arctan\left(0.5{\sqrt{2}}\mathrm{z}\right)\biggr]\;,\quad D_2(\mathrm{z})={\rm exp}\biggl[(\lambda_{1}-4\lambda_{2})\mathrm{z}+\lambda_{2}(1+\mathrm{z})^{2}\biggr]\;,\nonumber\\
F_{0}(\tilde{x})&=\biggl[\lambda_{0}+\lambda_{1}\tilde{x}+\lambda_{2}(2\tilde{x}^{2}-1)\biggr]
{|1+0.5z_{max}(1+\tilde{x})|}^{-(\lambda_{0}-\lambda_{1}+\lambda_{2}+3\omega_{2}+1)}\,,\quad F_{1}(\tilde{x})={\rm exp}\biggl[0.5z_{max}(1+\tilde{x})(4\lambda_{2}-\lambda_{1})\biggr]R_{2d}\;,\nonumber\\
F_2(\tilde{x})&=\frac{{\rm exp}\biggl[-\sqrt{2}\biggl(\omega_{0}+2\omega_{1}-5\omega_{2}\biggr)\arctan\biggl(0.25\sqrt{2}z_{max}(1+\tilde{x})\biggr)\biggr]}{\biggl[\left(\tilde{x}^{2}+2\right)(0.5z_{max})^{2}+2\left(1+z_{max}\right)\biggr]^{(\omega_{1}-\omega_{0}+5\omega_{2})}{\biggl[1+0.5z_{max}(1+\tilde{x})\biggr]}^{2(\omega_{o}-\omega_{1}+\omega_{2})}}\,,\nonumber\\
A_{0}&=\frac{8}{(2+z_{max})^{2}}\biggl[\omega_{0}-\omega_{1}+\omega_{2}+8\omega_{2}(z_{max}^{-2}+z_{max}^{-1})-2\omega_{1}z_{max}^{-1}\biggr]\;,\nonumber\\
A_{1}&=\frac{4}{(2+z_{max})^{2}}\biggl[\omega_{1}-\omega_{0}+5\omega_{2}+16\omega_{2}(z_{max}^{-2}+z_{max}^{-1})+2\omega_{1}z_{max}^{-1}\biggr]\;,\nonumber\\
A_{2}&=\frac{4}{(2+z_{max})^{2}}\biggl[\omega_{0}+2\omega_{1}-5\omega_{2}-16\omega_{2}(z_{max}^{-2}+z_{max}^{-1})+4\omega_{1}z_{max}^{-1}\biggr].
\end{align}
\end{widetext}
Eqs. (\ref{Interaction})-(\ref{ratio}) show that from simple arguments based on the evolution of $\mathrm{R}$, one can find an appropriated restriction for the coupling $\bar{Q}$ between the dark components of the 
universe \cite{Campo-Herrera2015}.\\ 
On the other hand, the coincidence problem can be alleviated, if we impose that $\bar{Q}\geq 0$ in Eq. (\ref{FormQ}). From here, and using Eq. (\ref{ratio}) at present time, we find
\begin{equation}\label{alleviate}
\biggl|\left(\frac{{\bar{\rm I}}_{\rm Q,0}(1+\mathrm{R_{0}})}{\omega_{0}}+3\right)\mathrm{R_{0}}\omega_{0}\biggr|\leq 3\mathrm{R_{0}}\;. 
\end{equation}
This result means that the slope of $\mathrm{R}$ at $\mathrm{z}=0$ is more gentle than that found in the $\Lambda$CDM model \cite{Campo-Herrera2015}.
\subsection{Crossing of ${\bar{\rm I}}_{\rm Q}=0$ line with a coupling.}\label{crossingIq}
From Eqs. (\ref{Interaction}) and (\ref{Chebyshev1}), we note that exist real values of $\mathrm{z}$ that leads to ${\bar{\rm I}}_{\rm Q}(\mathrm{z_{crossing}})=0$, which 
are called the redshift crossing points, $\mathrm{{z}_{crossing}}$
\begin{equation}
\label{crossIq} 2\lambda_{2}{\mathrm{{z}^{2}_{crossing}}}+\lambda_{1}{\mathrm{{z}_{crossing}}}+(\lambda_{0}-\lambda_{2})=0.
\end{equation}
Then, the solution of Eq. (\ref{crossIq}) is given by
\begin{equation}\label{ZcrossIq} 
\mathrm{{z}_{crossing}}=-\frac{\lambda_{1}}{4\lambda_{2}}\pm\sqrt{\left(\frac{\lambda_{1}}{4\lambda_{2}}\right)^{2}-\left(\frac{\lambda_{0}-\lambda_{2}}{2\lambda_{2}}\right)}.
\end{equation}
This result depends of the choice for ${\bar{\rm I}}_{\rm Q}$. However, the only possibility for a crossing happends when
\begin{equation}\label{dIqdznull2} 
\frac{\mathrm{d}{\bar{\rm I}}_{\rm Q}}{\mathrm{dz}}|_{\mathrm{z}_{crossing}}=\pm\frac{1}{4\lambda_{2}}\sqrt{{\lambda_{1}}^{2}-8\lambda_{2}\left(\lambda_{0}-\lambda_{2}\right)}\neq 0\,,\\
\end{equation}
From Eq. (\ref{dIqdznull2}), we impose the following restraint to gua-rantee real values in ${\lambda_{1}}$
\begin{eqnarray}
\label{dIqdznull6} 
\mid \lambda_{1} \mid &\geq& \sqrt{8\lambda_{2}\left(\lambda_{0}-\lambda_{2}\right)}\rightarrow \left(\lambda_{2}\geq 0\right)\,\cap\,\left(\lambda_{0}\geq\lambda_{2}\right)\nonumber\\
&&\cup\left(\lambda_{2}\leq 0\right)\,\cap\,\left(\lambda_{0}\leq\lambda_{2}\right)\,.
\end{eqnarray}
In general, if $\lambda_{0}$ and $\lambda_{2}$ are both positive or both negative, 
then $\mathrm{d}{\bar{\rm I}}_{\rm Q}/\mathrm{dz}$ could be positive or negative. Moreover, $\mathrm{d}{\bar{\rm I}}_{\rm Q}/\mathrm{dz}$ may be zero when $\lambda_{0}$, 
$\lambda_{1}$, and $\lambda_{2}$ are all zero (i.e. uncoupled DE models) or when $\mid \lambda_{1} \mid=\sqrt{8\lambda_{2}\left(\lambda_{0}-\lambda_{2}\right)}$. From here, 
we can describe the sign of $\bar{Q}$ (${\bar{\rm I}}_{\rm Q}$).
\section{Perturbed equations of motion.}\label{structure}
Let us consider a spatially flat universe with scalar perturbations about the background. The perturbed line element in the Newtonian 
gauge is given by \cite{valiviita2008,Jackson2009,He2009,Xu2011,He2011,Clemson2012}
\begin{equation}\label{metricFRW} 
{\rm ds}^{2}={\mathrm{a}}^{2}(\tau)\biggl[-\left(1+2\phi\right){\rm d}{\tau}^{2}+\left(1-2\psi\right)\delta_{ij}d{x}^{i}d{x}^{j}\biggr]\,,
\end{equation}
where $\tau$ is the conformal time, $\phi$ and $\psi$ are gravitational potentials, and ``$\mathrm{a}$'' is the scale factor. 
The four-velocity of fluid A is 
\begin{equation}\label{fourvelocity}
U^{\mu}_{A}=\mathrm{a}^{-1}(1-\phi,\partial^{i}v_{A})\,,\quad U_{\mu}^{A}=\mathrm{a}(-1-\phi,\partial_{i}v_{A})\,,
\end{equation}
where $v_{A}$ is the peculiar velocity potential, and $\theta_{A}=-k^{2}v_{A}$ is velocity perturbation in Fourier space. The energy-momentum tensor for each A is 
\cite{valiviita2008,Jackson2009,He2009,Xu2011,He2011,Clemson2012}
\begin{equation}\label{A_Energy_momentum_tensor}
{T}^{\mu}_{A\nu}=(\mathrm{\bar{\rho}_{A}}+{\bar{P}}_{A}){U^{\mu}_{A}}{U_{\nu}^{A}}+{g}^{\mu}_{\nu}{\bar{P}}_{A}\,,
\end{equation}
where the density $\mathrm{\rho_{A}}=\mathrm{\bar{\rho}_{A}}+\delta\rho_{A}$ and the pressure $P_{A}=\bar{P}_{A}+\delta P_{A}$. 
Then, the total energy-momentum tensor is ${T}^{\mu}_{\nu}=\sum_{A}^{}{T}^{\mu}_{A\nu}$ with $\rho=\sum_{A}^{}{\bar{\rho}}_{A}$ and $P=\sum_{A}^{}{\bar{P}}_{A}$).\\
Each fluid A satisfies the following energy-momentum balance equation \cite{valiviita2008,Jackson2009,He2009,Xu2011,He2011,Clemson2012}
\begin{equation}\label{Energy_momentum_transferA} 
\nabla_{\nu}{T_{A}}^{\mu\nu}=Q_{A}^{\mu}\,,\qquad \sum_{A}Q_{A}^{\mu}=0\,,
\end{equation}
where the four-vector $Q_{A}^{\mu}$ governs the energy-momentum exchange between the dark components and satisfies 
$Q_{DE}^{\mu}=-Q_{DM}^{\mu}$ \cite{valiviita2008,Clemson2012}.\\
A general $Q_{A}^{\mu}$ can be split relative to the total four-velocity as \cite{valiviita2008,Jackson2009,He2009,Xu2011,He2011,Clemson2012}
\begin{equation} \label{decomposeQA}
Q_{A}^{\mu}={Q}_{A}U^{\mu}_{A}+F^{\mu}_{A},\quad Q_{A}=\bar{Q}_{A}+\delta Q_{A},\quad U_{\mu}^{A}F^{\mu}_{A}=0,
\end{equation}
where ${Q}_{A}$ is the energy density transfer relative to $U^{\mu}_{A}$ and $F^{\mu}_{A}={\mathrm{a}}^{-1}(0,\partial^{i}f_{A})$ is the momentum density transfer rate, 
relative to $U^{\mu}_{A}$. Here, $f_{A}$ is a momentum transfer potential. We choose each $U^{\mu}_{A}$ and the total $U^{\mu}$ as 
\cite{Clemson2012}
\begin{equation}\label{Totaltensor}
{T}^{\mu}_{A\nu}U^{\nu}_{A}=-\mathrm{{\bar{\rho}}_{A}}U^{\nu}_{A}\,,\qquad  {T}^{\mu}_{\nu}U^{\nu}=-\mathrm{\bar{\rho}}U^{\mu}\,,
\end{equation}
Thus, the total energy-frame is defined as
\begin{equation}\label{velocitypotential}
(p+\rho)v = \sum_{A}(\mathrm{\bar{\rho}_{A}}+\mathrm{\bar{p}_{A}})v_{A}\,,
\end{equation}
where $v$ is the total energy-frame velocity potential. From Eqs. (\ref{fourvelocity}) and (\ref{decomposeQA}) obtain
\begin{equation}\label{componentF}
Q^{A}_{0}=-\mathrm{a}\left[\bar{Q}_{A}(1+\phi)+\delta{Q_{A}}\right]\,,\quad Q^{A}_{i}=\mathrm{a}\partial_{k}\left[f_{A}+\bar{Q}_{A}v\right]\,.  
\end{equation}
The perturbed energy transfer $Q^{A}_{0}$ includes a metric perturbation term ${\bar{Q}}_{A}\phi$ and a perturbation $\delta Q_{A}$. In addition, we stress that 
the perturbed momentum transfer $Q^{A}_{i}$ is made up of two parts: the momentum transfer potential ${\bar{Q}}_{A}v$ and $f_{A}$.\\
On the other hand, the physical sound-speed $c_{sA}$ of a fluid or scalar field $A$ is defined by $c_{sA}^{2}\equiv \delta{{P}_{A}}/\delta{\rho}_{A}\mid_{rf}$ in the $A$ rest-frame ($rf$), and the adiabatic 
sound-speed $c_{sa}$ can be defined as  $c_{\mathrm{a}A}^{2}\equiv {{P'}_{A}}/{{\rho'}_{A}}=\mathrm{\omega_{A}}+(\omega'_{A}/{\bar{\rho'}_{A}})\mathrm{\bar{\rho}_{A}}$ \cite{valiviita2008}.
For the adiabatic DM fluid, $c^{2}_{sDM}=c^{2}_{\mathrm{a}DM}=\omega_{DM}=0$. By contrast, the DE fluid is non-adiabatic and to avoid any unphysical instability, $c^{2}_{sDE}$ 
should be taken as a real and positive parameter. A common choice (and the one we make here) is to take $c^{2}_{sDE}=1$ \cite{valiviita2008,Clemson2012}.\\ 
Defining the density contrast as $\delta_{A}\equiv\delta\rho_{A}/\mathrm{{\bar{\rho}}_{A}}$, we can find equations for the density perturbations $\delta_{A}$ and 
the velocity perturbations $\theta_{A}$ \cite{valiviita2008,Clemson2012},
\begin{align}
\label{contrast1}
{\delta'}_{A}&=-3\bar{\mathcal{H}}(c^{2}_{sA}-\mathrm{\omega_{A}})\delta_{A}-9\bar{\mathcal{H}}^{2}(1+\mathrm{\omega_{A}})(c_{sA}^{2}-c_{aA}^{2})\frac{\theta_{A}}{\mathrm{k}^{2}} \nonumber\\
&-(1+\mathrm{{\omega}_{A}})\theta_{A}+3(1+\mathrm{{\omega}_{A}}){\psi'}_{A}+\frac{\mathrm{a}}{\mathrm{{\bar{\rho}}_{A}}}(-{\bar{Q}}_{A}\delta_{A}+\delta{Q}_{A})\nonumber\\
&+\frac{\mathrm{a}{\bar{Q}}_{A}}{\mathrm{{\bar{\rho}}_{A}}}\left[\phi+3\bar{\mathcal{H}}(c_{sA}^{2}-c_{aA}^{2})\frac{\theta_{A}}{\mathrm{k^{2}}}\right]\,,\\
\label{euler1} 
{{\theta'}_{A}}&=-\bar{\mathcal{H}}(1-3{c^{2}_{sA}}){{\theta}_{A}}+\frac{c^{2}_{sA}}{(1+\mathrm{{\omega}_{A}})}{\mathrm{k}^{2}}{{\delta}_{A}}-\frac{\mathrm{a}}{(1+\mathrm{{\omega}_{A}})
{\bar{\rho}_{A}}}\mathrm{k^{2}}f_{A}\nonumber\\
&+\frac{\mathrm{a}{\bar{Q}}_{A}}{(1+\mathrm{{\omega}_{A}}){\bar{\rho}_{A}}}\left[\theta-(1+c^{2}_{sA})\theta_{A}\right]+\mathrm{k^{2}}\phi\,,
\end{align}
The curvature perturbations on constant-$\rho_{A}$ surfaces and the total curvature perturbation are given by
\begin{equation}\label{curvature}
\zeta_{A}=-\psi-\bar{\mathcal{H}}\frac{{\delta_{A}}}{\bar{\rho'}_{A}}\,,\qquad \zeta=\sum_{A}\frac{{\bar{\rho'}_{A}}}{\mathrm{\bar{\rho}}}\zeta_{A}\,,
\end{equation}
Then, we need to specify a covariant form of $Q_{A}^{\mu}$ and $f_{A}$ in the dark sectors \cite{valiviita2008,Clemson2012}.
For $f_{A}$, the simplest physical choice is that there is no momentum transfer in the rest-frame of either DM or DE \cite{valiviita2008}. This leads to 
two cases 
\begin{eqnarray}
\label{CovariantQ1} 
{Q}_{DE}^{\mu}={\bar{Q}}_{DE}U^{\mu}_{DM}&=&-{Q}_{DM}^{\mu}\,,\qquad Q^{\mu}\parallel U^{\mu}_{DM}\,,\\
\label{CovariantQ2}
{Q}_{DE}^{\mu}={\bar{Q}}_{DE}U^{\mu}_{DE}&=&-{Q}_{DM}^{\mu}\,,\qquad Q^{\mu}\parallel U^{\mu}_{DE}\,.
\end{eqnarray}
Using Eqs. (\ref{fourvelocity}), (\ref{CovariantQ1}), (\ref{CovariantQ2}) and (\ref{componentF}), finds
\begin{equation}\label{totalconservation}
\sum_{A}\delta Q_{A}=\sum_{A}f_{A}=0\,,\qquad A=DM,\,\,DE\,.
\end{equation}
According to \cite{valiviita2008,Clemson2012}, $f_{A}$ can be
\begin{eqnarray}
\label{potential1} 
f_{DM}&=&\frac{{\bar{Q}}_{DM}}{k^{2}}(\theta-\theta_{DM})=-f_{DE}\,,\quad Q^{\mu}\parallel U^{\mu}_{DM}\,,\\
\label{potential2}
f_{DE}&=&\frac{{\bar{Q}}_{DE}}{k^{2}}(\theta-\theta_{DE})=-f_{DM}\,,\quad Q^{\mu}\parallel U^{\mu}_{DE}\,,
\end{eqnarray}
For both cases $Q^{\mu} \parallel U^{\mu}_{DM}$ or $Q^{\mu} \parallel U^{\mu}_{DE}$ cases, Eq. (\ref{contrast1}) does not change, but Eq. (\ref{euler1}) is different in both cases.\\
In this article, we focus only on the $Q^{\mu} \parallel U^{\mu}_{DM}$ case. Besides, assuming that $\bar{Q}$ depends only on the cosmic time through the 
global expansion rate, then a possible choice can be
\begin{equation}\label{IgualHubble}
 \delta{\bar{\mathcal{H}}}=0\,.
\end{equation}
From Eqs. (\ref{Interaction}), (\ref{decomposeQA}) and (\ref{IgualHubble}), we have
\begin{equation}\label{FDM}
Q_{DM}^{\mu}=Q_{DM}U_{DM}^{\mu}=(\bar{Q}_{DM}+\delta Q_{DM})U_{DM}^{\mu}\,.
\end{equation}
For convenience, we impose that ${{\delta{\rm I}}_{\rm Q}}\ll{\delta_{DM}}$, so 
\begin{equation}\label{parts1_Exchange_energy}
{\delta Q}_{DM}=-{\bar{\mathcal{H}}}{\bar{\rm I}}_{\rm Q}\mathrm{{\bar{\rho}}_{DM}}{\delta}_{DM}\;.
\end{equation}
In a forthcoming article we will extend our study, by considering other relations between ${{\delta{\rm I}}_{\rm Q}}$, ${\delta_{DM}}$ and $\delta{H}$. It is beyond the scope of the present paper.\\ 
Considering that $c^{2}_{sDM}=0$, $c^{2}_{sDE}=1$ and using the above Eqs into Eqs. (\ref{contrast1}) and 
(\ref{euler1}), we find \cite{valiviita2008,Mas}: 
\subsection{Dark sector}\label{DarkEqs}
\begin{align}\label{contrastDE}
{\delta'}_{DE}&=\bar{\mathcal{H}}(\mathrm{R}{\bar{\rm I}}_{\rm Q}-3-3\mathrm{\omega_{DE}})\delta_{DE}-\frac{\bar{\mathcal{H}}^{2}\theta_{DE}}{\mathrm{k^{2}}}\biggl[9(1-\mathrm{{\omega}_{DE}^{2}})\nonumber\\
&+\frac{3\omega'_{DE}}{\bar{\mathcal{H}}}+3\mathrm{R}{\bar{\rm I}}_{\rm Q}(1-\mathrm{\omega_{DE}})+\frac{\mathrm{k^{2}}{(1+\mathrm{\omega_{DE}})}}{\bar{\mathcal{H}}^{2}}\biggr]\nonumber\\
&+3(1+\mathrm{\omega_{DE}})\psi'-\bar{\mathcal{H}}\mathrm{R}{\bar{\rm I}}_{\rm Q}\phi-\bar{\mathcal{H}}\mathrm{R}{\bar{\rm I}}_{\rm Q}\delta_{DM}\,,\\
\label{eulerDE}
{\theta'}_{DE}&=2{{\theta}_{DE}}\bar{\mathcal{H}}\left(1+\frac{\mathrm{R}{\bar{\rm I}}_{\rm Q}}{(1+\mathrm{\omega_{DE}})}\right)+\frac{\mathrm{k^{2}}{\delta}_{DE}}{(1+\mathrm{\omega_{DE}})}+\nonumber\\
&+\mathrm{k^{2}}\phi+\frac{\bar{\mathcal{H}}\mathrm{R}{\bar{\rm I}}_{\rm Q}{\theta}_{DM}}{(1+\mathrm{\omega_{DE}})}\,,\\
\label{contrastDM}
{\delta'}_{DM}&=-\theta_{DM}+\bar{\mathcal{H}}{\bar{\rm I}}_{\rm Q}\phi+3\psi'\,,\\
\label{eulerDM} 
{\theta'}_{DM}&=-\bar{\mathcal{H}}\theta_{DM}+k^{2}\phi\,.
\end{align}
\subsection{Photon-baryon and Neutrino sectors}\label{PhotonBaryonNeutrinoEqs}
\begin{align}\label{contrastgb}
\delta'_{\gamma}&=-\frac{4}{3}\theta_{\gamma}+4\psi'\,,\quad \delta'_{b}=-\theta_{b}+3\psi'\,,\quad \delta'_{\nu}=-\frac{4}{3}\theta_{\nu}+4\psi'\,,\nonumber\\
{\theta'}_{\gamma}&=\frac{1}{4}\mathrm{k}^{2}\delta_{\gamma}+\mathrm{k}^{2}\phi\,,\quad {\theta'}_{b}=-\bar{\mathcal{H}}\theta_{b}+c^{2}_{sb}\mathrm{k}^{2}\delta_{b}+\mathrm{k}^{2}\phi\,,\nonumber\\
{\theta'}_{\nu}&=\frac{1}{4}\mathrm{k}^{2}\delta_{\nu}+\mathrm{k}^{2}\phi-\mathrm{k}^{2}\sigma_{\nu}\,,\quad \sigma'_{\nu}=\frac{4}{15}{\theta}_{\nu}\,,
\end{align}
Eq. (\ref{eulerDE}), shows an instability when $\mathrm{\omega_{DE}}\rightarrow -1$. Thus, we must exclude this value. 
\subsection{Radiation dominated era at super-horizon scales}\label{Radiation_era}
The following properties will simplify the solution of the perturbation equations for $\delta$ and $\theta$. Here,\\
\textbullet\, All scales of interest are outside the horizon, $\mathrm{k}\tau\ll1$.\\
\qquad This allows us to drop some of the terms with $\mathrm{k}$.\\ 
\textbullet\,$\mathrm{\bar{\rho}_{\gamma}},\,\mathrm{\bar{\rho}_{\nu}}\gg \mathrm{\bar{\rho}_{DM}},\,\mathrm{\bar{\rho}_{b}},\,\mathrm{\bar{\rho}_{DE}}$,$\hspace{0.2cm}$ $\omega_{\gamma}=\omega_{\nu}=1/3$\,,$\hspace{0.2cm}$ $\bar{\mathcal{H}}=1/\tau$.\\ 
\textbullet\, Baryons and photons are tightly coupled whose velocity perturbations are equal. Moreover, the photon distribution is isotropic.\\
\textbullet\, We assume massless neutrinos.\\
Thus, the perturbed Einstein equations become 
\begin{eqnarray}
\label{1Einstein}3\bar{\mathcal{H}}\psi'+3\bar{\mathcal{H}}^{2}\phi+\mathrm{k^{2}}\psi=-\frac{3}{2}\bar{\mathcal{H}}^{2}{\delta}\,,\qquad\qquad\qquad\qquad\qquad\\
\label{2Einstein}\mathrm{k^{2}}(\phi'+\bar{\mathcal{H}}\phi)=\frac{3}{2}\bar{\mathcal{H}}^{2}(1+\mathrm{\omega})\theta\,,\qquad\qquad\qquad\qquad\qquad\quad\\
\label{3Einstein}\phi''+2\bar{\mathcal{H}}\phi'-{\bar{\mathcal{H}}}^{2}\phi+\bar{\mathcal{H}}\phi'+\frac{\mathrm{k^{2}}}{3}(\psi-\phi)=\frac{3\bar{\mathcal{H}}^{2}}{2}\frac{\delta\,P}{\bar{\rho}}\,,\quad\quad\\
\label{4Einstein}(\psi-\phi)=8{\pi}G\pi_{\nu}\,.\qquad\qquad\qquad\qquad\qquad\qquad\qquad\qquad
\end{eqnarray}
From these Einstein equations and setting the following adiabatic initial conditions \cite{valiviita2008} 
\begin{eqnarray}
\phi&=&constant\,,\quad \psi=(1+\frac{2}{5}R_{\nu})\phi\,,\quad R_{\nu}=\,\frac{\mathrm{\bar{\rho}_{\nu}}}{\mathrm{\bar{\rho}_{\nu}}+\mathrm{\bar{\rho}_{\gamma}}}\,,\nonumber\\
\delta_{\gamma}&=&\delta_{\nu}=\frac{4}{3}\delta_{b}=\frac{4}{3}\delta_{DM}=4\delta_{DE}=-2\phi,\,\sigma_{\nu}=\frac{1}{15}{(\mathrm{k}\tau)}^{2}\phi,\nonumber\\
\theta_{\gamma}&=&\theta_{\nu}=\theta_{b}=\theta_{DM}=\theta_{DE}=\frac{\mathrm{k}^{2}\phi\tau}{2},
\end{eqnarray}
we will solve numerically Eqs. (\ref{contrastDE})-(\ref{contrastgb}) in the early universe with $\bar{\mathcal{H}}=\mathrm{a\bar{H}}$ and $\mathrm{a}=\mathrm{H_{0}}\sqrt{\Omega_{r,0}}\tau$. 
We have also fixed $\Omega_{\gamma,0}=2.469\times10^{-5}h^{-2}$, $\Omega_{b,0}=0.02230h^{-2}$, and $\Omega_{r,0}=\Omega_{\gamma,0}(1+0.2271N_{eff})$, where 
$N_{eff}$ represents the effective number of neutrino species (here, $\Omega_{b,0}$, $\Omega_{r,0}$ and the standard value, $N_{eff}=3.04\pm0.18$ were chosen from 
Table $4$ in \cite{Planck2015}).\\
Substituting Eq. (\ref{eulerDE}) into Eq. (\ref{contrastDE}) and using Eq. (\ref{approach}), we find a second order differential equation for $\delta_{DE}$ du-ring the 
radiation dominated era
\begin{equation}\label{SecondDE}
{\delta''}_{DE}=-3\left(\frac{1+3\omega_{DE}}{1+\omega_{DE}}\right)\frac{{\delta'}_{DE}}{\tau}-\frac{6}{\tau^{2}}{\delta}_{DE}\,.
\end{equation}
This equation can be solved in the form of a power law ${\delta}_{DE}=A_{r}{\tau}^{r}$, where the coefficients $A_{r}$ and the indices $\mathrm{r}$ are real numbers.
The index $\mathrm{r}$ is given by
\begin{equation}\label{indices}
\mathrm{r}=\mathrm{r_{\pm}}=-\left(\frac{1+4\mathrm{\omega_{DE}}}{1+\mathrm{\omega_{DE}}}\right)\pm \frac{\sqrt{10\mathrm{\omega_{DE}^{2}}-4\mathrm{\omega_{DE}}-5}}{1+\mathrm{\omega_{DE}}}\,. 
\end{equation}
This analyical approximation does not consider the value of ${\bar{\rm{I}}}_{\rm{Q}}$, and thus, it does not give a real description of the behaviour of the DE perturbation in the early universe. 
For this reason, we need to solve numerically Eqs. (\ref{contrastDE}) and (\ref{eulerDE}), respectively.
In this article, we focus on the analysis of the different effects of including the reconstructions of $\bar{Q}$ and $\mathrm{\omega_{DE}}$ on the stability of the DE perturbations in the early universe and also 
on the evolution of the linear growth rate of DM density perturbation, ${\delta}_{DM}$ at late times.\\
\subsection{Matter dominated era at sub-horizon scales}\label{Matter_era}
Combining Eqs. (\ref{contrastDM}) and (\ref{eulerDM}), and using the Eq. (\ref{1Einstein}) at small scales (Poisson equation), we can eliminate $\theta_{DM}$ and obtain a 
second-order equation for the DM density perturbations
\begin{equation}\label{growth_ratez} 
\frac{\mathrm{d}^{2}{\delta}_{DM}}{\mathrm{d}{\mathrm{z}}^{2}}=-\frac{(1+3\mathrm{\omega_{DE}}\mathrm{\Omega_{DE}})}{2(1+\mathrm{z})}
\frac{\mathrm{d}{\delta}_{DM}}{\mathrm{dz}}+\frac{3\mathrm{\bar{\Omega}_{DM}}\delta_{DM}}{2(1+\mathrm{z})^{2}}\,.
\end{equation}
This Eq. has the same form as that for the uncoupled models (standard DM growth equation). Nevertheless, for the coupled models $\mathrm{\bar{\rho}_{DM}}$, $\mathrm{\bar{\rho}_{DE}}$ 
and $\mathrm{\omega_{DE}}$ evolve significantly different, affecting ${\delta}_{DM}$. In the matter era, Eq. (\ref{growth_ratez}) has a growing-mode solution $\delta_{DM}\,\alpha\,\mathrm{a}$. 
Furthermore, It can be solved numerically, considering that $\delta(\mathrm{z}=0)=1$ and $\delta^{'}(\mathrm{z}=0)=-\Omega_{DM}(\mathrm{z}=0)^{\gamma(\mathrm{z}=0)}$. Here, $\gamma$ is 
a some unknown function of $\mathrm{z}$ so-called the growth index of the linear matter fluctuations. In the linear regime, it is convenient to define,
\begin{equation}\label{f} 
\mathrm{f}\equiv\frac{\mathrm{d}\ln\delta}{\mathrm{d}{\ln \mathrm{a}}}=-(1+\mathrm{z})\frac{\mathrm{d}{\ln}\delta}{\mathrm{dz}}\,, 
\end{equation}
called the growth factor of DM density perturbations. Then, Eq. (\ref{growth_ratez}) can be rewritten in function of $\mathrm{f}$ as 
\begin{equation}\label{growth_factor} 
\frac{\mathrm{df}}{\mathrm{dz}}=\frac{\mathrm{f^{2}}}{(1+\mathrm{z})}+\mathrm{f}\frac{(1-3\mathrm{\omega_{DE}}\mathrm{\Omega_{DE}})}{2(1+\mathrm{z})}-\frac{3\mathrm{\bar{\Omega}_{DM}}}{2(1+\mathrm{z})^{2}}\,.
\end{equation}
The above equation can be solved numerically, taking into account the condition $f(0)={\Omega_{DM}(\mathrm{z}=0)}^{\gamma(\mathrm{z}=0)}$.\\ 
On the other hand, the root-mean-square amplitude of matter density perturbations within a sphere of radius $8\,Mpc h^{-1}$ (being $h$ the dimensionless Hubble parameter) 
is denoted as $\mathrm{\sigma_{8}(z)}$ and its evolution is represented by
\begin{equation}\label{Sigma8}
\mathrm{\sigma_{8}(z)}=g(\mathrm{z})\mathrm{\sigma_{8}(z=0)}\,, \qquad g(\mathrm{z}) \equiv \frac{\delta_{DM}(\mathrm{z})}{\delta_{DM}(0)}\,.
\end{equation}
where $\mathrm{\sigma_{8}(z=0)}$ and ${\delta_{DM}(0)}$ are the normalizations to unity of $\mathrm{\sigma_{8}(z)}$ and $\delta_{DM}(\mathrm{z})$ today, respectively. 
Thus, the functions f y $\mathrm{\sigma_{8}}$ can be combined to obtain f$\mathrm{\sigma_{8}}$ at different redshifts. From here, we obtain
\begin{equation}\label{fs8}
\mathrm{f(z)}\mathrm{\sigma_{8}(z)}=\mathrm{f(z)}g(\mathrm{z})\mathrm{\sigma_{8}(z=0)}.
\end{equation}
The measurements of $\mathrm{f\sigma_{8}}$ will be important to constrain different cosmological models.
\subsection{DE dominated era at sub-horizon scales}\label{DE_era}
Here, we consider the evolution of DM structure formation when DE dominates in the universe and ${\bar{\rm I}}_{\rm Q}$ is very small at late times. From Eqs. 
(\ref{EDE}) and (\ref{hubble}), we have
\begin{equation}\label{H_DE}
\bar{\mathcal{H}}=\frac{2}{\left(6\omega+{\lambda}_{1}-{\lambda}_{0}-{\lambda}_{2}+1\right)\eta}
\end{equation}
where $\eta=\tau-\tau_{\infty}$, $\tau_{\infty}$ is a constant of integration and represents the radius of the Sitter event horizon in the uncoupled models with a 
cosmological constant. Besides, when $\omega=\omega_{0}+\omega_{1}$ we are in the XCPL model and with $\omega=\omega_{2}$ we are in the DR model.\\
Using Eqs. (\ref{growth_ratez}) and (\ref{growth_factor}), the DM density perturbations and DM growth factor can then be written as
\begin{eqnarray}
\label{growth_DE} 
\frac{\mathrm{d}^{2}{\delta}_{DM}}{\mathrm{d}z^{2}}&=&-\frac{(1+3\mathrm{\omega_{DE}}\mathrm{\Omega_{DE}})}{2(1+z)}\frac{\mathrm{d}{\delta}_{DM}}{\mathrm{d}z}+\nonumber\\
&& +\frac{3\delta_{DM}}{2(1+z)^{2}}\left(1+\frac{2}{3}{\bar{\rm I}}_{\rm Q}-\frac{{\bar{\rm I}}_{\rm Q}}{3\mathrm{\omega_{DE}}}\right)\,,\\
\label{growthfactor_DE} 
\frac{\mathrm{df}}{\mathrm{dz}}&=&\frac{\mathrm{f}^{2}}{(1+\mathrm{z})}+\mathrm{f}\frac{(1-3\mathrm{\omega_{DE}}\mathrm{\Omega_{DE}})}{2(1+z)}+\nonumber\\
&& -\frac{3}{2(1+z)}\left(1+\frac{2}{3}{\bar{\rm I}}_{\rm Q}-\frac{{\bar{\rm I}}_{\rm Q}}{3\mathrm{\omega_{DE}}}\right)\,.
\end{eqnarray}
From these equations, we note that the term $(1+{2}/{3}{\bar{\rm I}}_{\rm Q}-{{\bar{\rm I}}_{\rm Q}}/{3\mathrm{\omega_{DE}}})$ reduces the value of the DM fluctuations, 
diminishing the concentrations of ${\delta}_{DM}$ and $\mathrm{f}$, respectively. Furthermore, they can be solved numerically, by using the above initial conditions.
\section{Current observational data.} \label{SectionAllTest}
In this section, we describe how we use the cosmological data currently available to test and put tighter constraints on the values of the cosmological parameters.
\subsection{Join Analysis Luminous data set (JLA).} \label{SNIa}
The SNe Ia data sample used in this work is the Join Analysis Luminous data set (JLA) \cite{Betoule2014} composed by $740$ SNe with hight-quality light curves. Here, JLA data
include samples from $\mathrm{z}<0.1$ to $0.2<\mathrm{z}<1.0$.\\
For the JLA data, the observed distance modulus of each SNe is modeled by
\begin{eqnarray}\label{muJLA}
{\mu}^{JLA}_{i}={m}^{*}_{B,i}+\alpha {x_{1,i}}-\beta C_{i}-M_{B}\,,
\end{eqnarray}
where and the parameters ${m}^{*}_{B}$, $x_{1}$ and $C$ describe the intrinsic variability in the luminosity of the SNe. Furthermore, the nuisance 
parameters $\alpha$, $\beta$, $M$ and $dM$ characterize the global properties of the light-curves of the SNe and are estimated simultaneously with the cosmological 
parameters of interest. Then, we defined $M_{B}$
\begin{equation}\label{Mb}
M_{B}=\left\{\begin{array}{cl}
M,\, &\mbox{$if$}\,\,\,\rm{M_{stellar}}\,\,\,<10^{10}\rm{M_{\bigodot}}\,,\\
M+dM,\, &\mbox{$if$}\,\,\,\rm{otherwise}\,,           
\end{array}\right.
\end{equation}
where $\rm{M_{stellar}}$ is the host galaxy stellar mass, and $\rm{M_{\bigodot}}$ is the solar mass. The details of building of the matrix $C_{\bf Betoule}$ can be found 
in \cite{Conley2011,Jonsson2010,Betoule2014}.\\
On the other hand, the theoretical distance modulus is 
\begin{equation}\label{mus}
{\mu}^{\rm{th}}(\mathrm{z},\mathbf{X}) \equiv 5{\log}_{10}\left[\frac{{D_{L}}(\mathrm{z},\mathbf{X})}{\rm{Mpc}}\right]+25\;,
\end{equation}
where ``$\rm{th}$'' denotes the theoretical prediction for a SNe at $\mathrm{z}$. The luminosity distance ${D_{L}}(\mathrm{z},\mathbf{X})$, is defined as
\begin{equation}\label{luminosity_distance1}
{D}_{L}(z_{hel},z_{CMB},\mathbf{X})=(1+z_{hel})c \int_{0}^{z_{CMB}}\frac{dz'}{\mathrm{H}(z',\mathbf{X})}\;,
\end{equation}
where $z_{hel}$ is the heliocentric redshift, $z_{CMB}$ is the CMB rest-frame redshift, ``$c=2.9999\times10^{5}km/s$'' is the speed of the light and $\mathbf{X}$ represents 
the cosmological parameters of the model. Thus, we rewrite ${\mu}^{{\rm th}}(\mathrm{z},\mathbf{X})$ as
\begin{eqnarray}\label{mus}
{\mu}^{{\rm th}}(z_{hel},z_{CMB},\mathbf{X})&=&5\log_{10}\biggl[(1+z_{hel}\int_{0}^{z_{CMB}}\frac{dz'}{E(z',\mathbf{X})})\biggr]\nonumber\\
&&+52.385606-5\log_{10}(\mathrm{H_{0}})\,.
\end{eqnarray}
Then, the ${\chi}^{2}$ distribution function for the JLA data is
\begin{equation}\label{X2JLA}
{\chi}_{\bf JLA}^{2}(\mathbf{X})=\left({\Delta{\mu}}_{i}\right)^{t}\left(C^{-1}_{\bf Betoule}\right)_{ij}\left({\Delta{\mu}}_{j}\right)\,,
\end{equation}
where ${\Delta \mu}_{i}={\mu}^{th}_{i}(\mathbf{X})-{\mu}^{JLA}_{i}$ is a column vector of $740$ entries of residuals between the theoretical and distance modulus, and 
$C^{-1}_{\bf Betoule}$ is the $740\times740$ covariance matrix for all the observed distance modulus reported in \cite{Betoule2014}.
\subsection{RSD data}\label{RSDdata}
RSD data are a compilation of measurements of the quantity $\mathrm{f\sigma_{8}}$ at different redshifts, and obtained in a model independent way. 
These data are apparent anisotropies (effects) of the galaxy distribution in redshift space due to the differences of the estimates between the redshifts observed distances 
and true distances. They are caused by the component along the line of sight (LOS) of the peculiar velocity of each of the galaxies 
(recessional speed) \cite{Jackson1972, Kaiser1987}.\\
In this work, we utilize the growth rate data collected by Mehrabi et al. (see Table in \cite{Mehrabi2015}). 
The standard $\chi^2$ for this data set is defined as \cite{Mehrabi2015}
\begin{equation}\label{X2RSD}
{\chi}^{2}_{RSD}(\mathbf{X}) \equiv \sum_{i=1}^{18}\frac{\left[\mathrm{f\sigma_{8}}^{\mathrm{th}}(\mathbf{X},\mathrm{z_{i}})-\mathrm{f\sigma_{8}}^{\mathrm{obs}}(\mathrm{z_{i}})\right]^{2}}{{\sigma}^{2}(\mathrm{z_{i}})}\;\;,
\end{equation}
where $\sigma(\mathrm{z_{i}})$ is the observed $1\sigma$ uncertainty, $\mathrm{f\sigma_{8}}^{\rm{th}}(\mathbf{X},\mathrm{z_{i}})$ and $\mathrm{f\sigma_{8}}^{\mathrm{obs}}(\mathrm{z_{i}})$ represent the theoretical and 
observational growth rate, respectively.
\begingroup
\squeezetable
\begin{table}
\centering
\begin{tabular}{| c | c | c | c |c | c | c | c |}
 \hline
$\mathrm{z}$ & $\mathrm{{f\sigma8}^{obs}}$ & $\sigma$ & Refs. &$\mathrm{z}$ & $\mathrm{{f\sigma8}^{obs}}$ & $\sigma$ & Refs.\\ 
 \hline
$0.020$ & $0.360$& $\pm0.0405$&\cite{Hudson2013} &$0.400$ & $0.419$ & $\pm0.041$ & \cite{Tojeiro2012}  \\
$0.067$ & $0.423$& $\pm0.055$ &\cite{Beutler2012}   &$0.410$ & $0.450$ & $\pm0.040$ & \cite{Blake2011}\\
$0.100$ & $0.370$& $\pm0.130$ &\cite{Feix2015}   &$0.500$ & $0.427$ & $\pm0.043$ & \cite{Tojeiro2012}\\
$0.170$ & $0.510$& $\pm0.060$ &\cite{Percival2004}   &$0.570$ & $0.427$ & $\pm0.066$ & \cite{Reid2012}\\      
$0.220$ & $0.420$& $\pm0.070$ &\cite{Blake2011}   &$0.600$ & $0.430$ & $\pm0.040$ & \cite{Blake2011}\\
$0.250$ & $0.351$& $\pm0.058$ &\cite{Samushia2012}   &$0.600$ & $0.433$ & $\pm0.067$ & \cite{Tojeiro2012}\\ 
$0.300$ & $0.407$& $\pm0.055$ &\cite{Tojeiro2012}   &$0.770$ & $0.490$ & $\pm0.180$ & \cite{Song2009,Guzzo2008}\\
$0.350$ & $0.440$& $\pm0.050$ &\cite{Song2009,Tegmark2006}  &$0.780$ & $0.380$ & $\pm0.040$ & \cite{Blake2011}\\
$0.370$ & $0.460$& $\pm0.038$ &\cite{Samushia2012}   &$0.800$ & $0.470$ & $\pm0.080$ & \cite{delaTorre2013}\\
\hline
\end{tabular}
\caption{Summary of RSD data set \cite{Hudson2013,Beutler2012,Feix2015,Percival2004,Song2009,Tegmark2006,Guzzo2008,
Samushia2012,Blake2011,Tojeiro2012,Reid2012,delaTorre2013}.}
 \label{tableRSD}
\end{table}
\endgroup
\subsection{BAO data sets}\label{BAO}
\subsubsection{$\bf{\rm{BAO}}$\,\,$\bf{\rm{I}}$\, data}\label{BAOI}
Here, we use a compilation of measurements of the distance ratios $\mathrm{d_{z}}$, obtained from different surveys \cite{Hinshaw2013,Beutler2011,Ross2015,Percival2010,
Kazin2010,Padmanabhan2012,Chuang2013a,Chuang2013b,Anderson2014a,Kazin2014,Debulac2015,FontRibera2014}.
Eisenstein et al. \cite{Eisenstein1998} and Percival et al. \cite{Percival2010} constructed an effective distance ratio $D_{v}(\mathrm{z})$ to encode the visual 
distortion of a spherical object due to the non-Euclidianity of a FRW spacetime,
\begin{equation}\label{Dv}
D_{v}(\mathrm{z},\mathbf{X})\equiv\frac{1}{\mathrm{H_{0}}}\left[(1+\mathrm{z})^{2}{D_{A}}^{2}(\mathrm{z})\frac{c\mathrm{z}}{E(\mathrm{z})}\right]^{1/3}\,,
\end{equation}
where $D_{A}(\mathrm{z})$ is the angular diameter distance given by
\begin{eqnarray} \label{DA}
D_{A}(\mathrm{z},\mathbf{X}) &\equiv& \frac{c}{(1+\mathrm{z})}{\int}^{\mathrm{z}}_{0}\frac{dz'}{\mathrm{H}(z',\mathbf{X})}\;.
\end{eqnarray}
The comoving sound horizon size is defined by
\begin{equation} \label{hsd} 
r_{s}(\mathrm{a})\equiv c\int^{\mathrm{a}}_{0}\frac{c_{s}(a')da'}{{a'}^{2}H(a')}\;\;,
\end{equation}
being $c_{s}(a)$ the sound speed of the photon-baryon fluid
\begin{equation} \label{vsd}
c_{s}^{2}(\mathrm{a}) \equiv \frac{\delta P}{\delta \rho}=\frac{1}{3}\left[\frac{1}{1+(3\Omega_{b}/4\Omega_{r})\mathrm{a}}\right]\;\;.
\end{equation}
Considering Eqs. (\ref{hsd}) and (\ref{vsd}) for $\mathrm{z}$, we have   
\begin{equation}\label{rs}
r_{s}(\mathrm{z})=\frac{c}{\sqrt 3}{\int}^{1/(1+\mathrm{z})}_{0}\frac{\mathrm{da}}{\mathrm{{a}^{2}}\mathrm{H(a)}\sqrt{1+(3\Omega_{b,0}/4\Omega_{\gamma,0})\mathrm{a}}}\;.
\end{equation}
The epoch in which the baryons were released from photons is denoted as, $z_{d}$, and can be determined by \cite{Eisenstein1998}:
\begin{equation}\label{zd}
\mathrm{z_{d}}=\frac{1291(\Omega_{M,0}h^{2})^{0.251}}{1+0.659(\Omega_{M,0}h^{2})^{0.828}}\left(1+b_{1}(\Omega_{b,0}h^{2})^{b_{2}}\right)\;,
\end{equation}
where $\Omega_{M,0}=\Omega_{DM,0}+\Omega_{b,0}$, and
\begin{eqnarray*} 
b_{1}&=&0.313(\Omega_{M,0}h^{2})^{-0.419}\left[1+ 0.607(\Omega_{M,0}h^{2})^{0.674}\right]\;,\\
b_{2}&=&0.238(\Omega_{M,0}h^{2})^{0.223}\;.
\end{eqnarray*}
The peak position of the BAO depends of the distance radios $\mathrm{d_{z}}$ at different redshifts, and listed in Table \ref{tableBAOI}.\\
\begin{equation} \label{dvalues}
\mathrm{d_{z}(\mathbf{X})}=\frac{r_{s}(\mathrm{z_{d}})}{D_{V}(\mathrm{z},\mathbf{X})}\;,
\end{equation}
where $r_{s}(\mathrm{z_{d}},\mathbf{X})$ is the comoving sound horizon size at the baryon drag epoch. From Table \ref{tableBAOI}, the $\chi^{2}$ becomes 
\begingroup
\squeezetable
\begin{table}
\centering
\begin{tabular}{| c | c | c | c |c | c |c | c |}
 \hline
 $\mathrm{z}$ & $\mathrm{d_{z}^{obs}}$ & $\sigma_{z}$& Refs. & $\mathrm{z}$ & $\mathrm{d_{z}^{obs}}$ & $\sigma$& Refs.\\
 \hline
$0.106$ & $0.3360$ &$\pm0.0150$ &\cite{Hinshaw2013,Beutler2011} &$0.350$ & $0.1161$ &$\pm0.0146$&\cite{Chuang2013a}\\
$0.150$ & $0.2232$ &$\pm0.0084$ &\cite{Ross2015} &$0.440$ & $0.0916$ &$\pm0.0071$&\cite{Blake2011}\\
$0.200$ & $0.1905$ &$\pm0.0061$ &\cite{Percival2010,Blake2011} &$0.570$ & $0.0739$ &$\pm0.0043$&\cite{Chuang2013b}\\
$0.275$ & $0.1390$ &$\pm0.0037$ &\cite{Percival2010} &$0.570$ & $0.0726$ &$\pm0.0014$&\cite{Anderson2014a}\\
$0.278$ & $0.1394$ &$\pm0.0049$ &\cite{Kazin2010} &$0.600$ & $0.0726$ &$\pm0.0034$&\cite{Blake2011}\\
$0.314$ & $0.1239$ &$\pm0.0033$ &\cite{Blake2011} &$0.730$ & $0.0592$ &$\pm0.0032$&\cite{Blake2011}\\
$0.320$ & $0.1181$ &$\pm0.0026$ &\cite{Anderson2014a}&$2.340$ & $0.0320$ &$\pm0.0021$&\cite{Debulac2015}\\
$0.350$ & $0.1097$ &$\pm0.0036$ &\cite{Percival2010,Blake2011} &$2.360$ & $0.0329$ &$\pm0.0017$&\cite{FontRibera2014}\\
$0.350$ & $0.1126$ &$\pm0.0022$ &\cite{Padmanabhan2012}        &$ $&$ $\\ 
 \hline
\end{tabular}
\caption{Summary of BAO data set \cite{Percival2010,Hinshaw2013,Beutler2011,Ross2015,Blake2011,Kazin2010,Padmanabhan2012,Chuang2013a,Chuang2013b,Anderson2014a,Debulac2015,
FontRibera2014}.}\label{tableBAOI}
\end{table}
\endgroup
\begin{equation}\label{X2BAOI}
\chi_{\bf{\bm{BAO}}\,\rm{I}}^{2}(\mathbf{X})=\sum_{i=1}^{17}\left(\frac{d_{z}^{\mathrm{th}}(\mathbf{X},\mathrm{z_{i}})-d_{z}^{\mathrm{obs}}(\mathbf{X},\mathrm{z_{i}})}{\sigma(\mathbf{X},\mathrm{z_{i}})}\right)^{2}\;\;.
\end{equation}
\subsubsection{$\bf{\rm{BAO}}$\,\,$\bf{\rm{II}}$\,data}\label{BAOII}
From BOSS DR $9$ CMASS sample, Chuang et al. in \cite{Chuang2013b} analyzed the shape of the monopole and quadrupole from the two-dimensional two-points 
correlation function $2$d$2$pCF of galaxies and measured simultaneously $\mathrm{H(z)}$, $D_{A}(\mathrm{z})$, $\Omega_{m}h^{2}$ and $\mathrm{f(z)\sigma_{8}(z)}$ at the effective redshift $\mathrm{z}=0.57$. 
From here, Chuang et al. defined ${\Delta A}_{i}=A^{\mathrm{th}}_{i}(\mathbf{X})-A^{\mathrm{obs}}_{i}$ as a column vector
\begin{equation}\label{bossdr9}
{\Delta A}_{i}=\left(\begin{array}{rl}
\mathrm{H(0.57)}-87.6\\
D_{A}(0.57)-1396\\
\Omega_{m}h^{2}(0.57)-0.126\\
\mathrm{f(0.57)\sigma_{8}(0.57)}-0.428\\
\end{array}\right)\;,
\end{equation}
Then, the $\chi^{2}$ function for the BAO $\rm{II}$ data is given by 
\begin{equation}\label{X2BAOII}
\chi_{\bf BAO\,\rm{II}}^{2}(\mathbf{X})=\left({\Delta A}_{i}\right)^{t}\left(C^{-1}_{\bf BAO\,\rm{II}}\right)_{ij}\left({\Delta A}_{j}\right),
\end{equation}
where the covariance matrix is listed in Eq. ($26$) of \cite{Chuang2013b}
\begin{equation}\label{BAOII}
C^{-1}_{\bf{\rm{BAO}}\,\bf{\rm{II}}}=\left(\begin{array}{lccr}
 +0.03850 \,\,-0.0011410 \,-13.53 \,-1.2710\\
 -0.001141 +0.0008662   \,+3.354 \,-0.3059\\
 -13.530  \,\,\,\,\,+3.3540\quad\,\,\,\,\,+19370\,-770.0\\
 -1.2710  \,\,\,\,\,-0.30590 \quad\,-770.0\,\,+411.3
\end{array} \right)\;.
\end{equation}
where ``t'' denotes its transpose.
\subsubsection{$\bf{\rm{BAO}}$\,\,$\bf{\rm{III}}$\, data}\label{BAOIII}
Using SDSS DR $7$ sample Hemantha et al \cite{Hemantha2014}, proposed a new method to constrain $\mathrm{\bar{H}}$ and $D_{A}$ simultaneously from the two-dimensional matter power 
spectrum $2$dMPS without assuming a DE model or a flat universe. They defined a column vector ${\Delta B}_{i}=B^{\mathrm{th}}_{i}(\mathbf{X})-B^{\mathrm{obs}}_{i}$ as
\begin{equation}
B^{\mathrm{th}}_{i}(\mathbf{X})-B^{\mathrm{obs}}_{i}=\left(\begin{array}{rl}                               
 \mathrm{H(0.35,\mathbf{X})}-81.3\\
 D_{A}(0.35,\mathbf{X})-1037.0\\
 \Omega_{M}h^{2}(0.35,\mathbf{X})-0.1268\\
\end{array}\right)\;.
\end{equation}
The covariance matrix for the set of parameters was
\begin{equation}\label{BAOIII}
C^{-1}_{\bf BAO\,\rm{III}}=\left(\begin{array}{lcr}           
+0.00007225 \, -0.169606 \,+0.01594328\\
-0.1696090  \,\,\,+1936.0\,\,\quad+67.030480 \\
+0.01594328 \, +67.03048\,+14.440\\
\end{array} \right)\;.
\end{equation}
The $\chi^{2}$ function for these data can be written as  
\begin{equation}\label{X2BAOIII}
\chi_{\bf{\rm{BAO}}\,\bf{\rm{III}}}^{2}(\mathbf{X})=\left({\Delta B}_{i}\right)^{t}\left(C^{-1}_{\bf BAO \rm{III}}\right)_{ij}\left({\Delta B}_{j}\right),
\end{equation}
where ``t'' denotes its transpose.
\subsubsection{$\bf{\rm{BAO}}$\,\,$\bf{\rm{IV}}$\, data}\label{BAOIV}
In order to measure the position of the clustering of galaxies, we need to convert angular positions and redshifts of galaxies into physical positions. It can be obtained, 
using a fiducial cosmological model. If there is significantly different from the real (true) cosmology, then this difference will induce any measured 
anisotropy, and could be used to constrain the true cosmology of the universe. It is also known as the AP test. This signal can be conveniently combined in a single 
parameter known as the $\mathrm{AP}$ distortion parameter $F_{AP}(\mathrm{z})$, defined \cite{Alcock1979} as
\begin{equation}\label{Fap}
F_{AP}(\mathrm{z})=(1+\mathrm{z})D_{A}(\mathrm{z})\left(\mathrm{H(z)}/c\right)\,. 
\end{equation}
Measuring this parameter we could break the degene-racy between $D_{A}$ and $\mathrm{\bar{H}}$ \cite{Seo2008}. Here, it is convenient to define the joint measurements of $\mathrm{d_{z}(z_{eff}})$, 
$F_{AP}(\mathrm{z_{eff}})$ and $\mathrm{f(z_{eff})\sigma_{8}(z_{eff})}$ in a vector $V$ evaluated at the effective redshift $\mathrm{z_{eff}}=0.57$ \cite{Anderson2014a, Battye2015, Samushia2014}
\begin{equation}\label{bossdr11}
{\Delta V}_{i}=V^{\mathrm{th}}_{i}(\mathbf{X})-V^{\mathrm{obs}}_{i}=\left(\begin{array}{rl}
\mathrm{d_{z}(z_{eff})}-13.880\\
F_{AP}(\mathrm{z_{eff}})-0.683\\
\mathrm{f(z_{eff})\sigma_{8}(z_{eff})}-0.422\\
\end{array}\right)\;,
\end{equation}
The $\chi^{2}$ function for this data set is fixed as
\begin{equation}\label{X2BAOIV}
\chi_{\bf BAO\,\rm{IV}}^{2}(\mathbf{X})=\left({\Delta V}_{i}\right)^{t}\left(C^{-1}_{\bf BAO\,\rm{IV}}\right)_{ij}\left({\Delta V}_{j}\right)\,,
\end{equation}
where the covariance matrix is listed in Eq. ($1.3$) of \cite{Battye2015}
\begin{equation}\label{BAOIV}
C^{-1}_{\bf{\rm{BAO}}\,\bf{\rm{IV}}}=\left(\begin{array}{lcr}           
+31.032  \,+77.773 \,-16.796\\
+77.773 \, +2687.7 \,-1475.9\\
-16.796 \, -1475.9 \,+1323.0\\
\end{array} \right)\;.
\end{equation}
Considering the Eqs. (\ref{X2BAOI}), (\ref{X2BAOII}), (\ref{X2BAOIII}) and (\ref{X2BAOIV}), we can construct the total $\chi_{\bf BAO}^{2}$ for all the BAO data sets
\begin{equation}\label{X2Total}
 {{{\rm{\bf{\chi}}^{2}}}}_{\bf BAO}=\chi_{\bf{BAO\,\rm{I}}}^{2}+\chi_{\bf{BAO\,\rm{II}}}^{2}+\chi_{\bf{BAO\,\rm{III}}}^{2}+\chi_{\bf{BAO\,\rm{IV}}}^{2}\,.
\end{equation}
\subsection{CMB data set} \label{CMB}
The JLA (SNe Ia) and BAO data sets contain information about the universe at low redshifts, we now include Planck $2015$ data \cite{Planck2015} to probe the
entire expansion history up to the last scattering surface. Here, the shift parameter ${\rm {\bf R}}$ is defined by \cite{Bond-Tegmark1997}
\begin{equation}\label{Shiftparameter}
{\rm {\bf R}}(z_{*},\mathbf{X})\equiv\sqrt{\Omega_{M,0}}{\int}^{z_{*}}_{0}\frac{d\tilde{y}}{E(\tilde{y})}\,,
\end{equation}
where $E(\tilde{y})$ is given by Eq. (\ref{hubble}). The redshift $z_{*}$ (the decoupling epoch of photons) is obtained using \cite{Hu-Sugiyama1996}
\begin{equation}\label{Redshift_decoupling}
{z}_{*}=1048\biggl[1+0.00124({\Omega}_{b,0}h^{2})^{-0.738}\biggr]\biggl[1+{g}_{1}({\Omega}_{M,0}h^{2})^{{g}_{2}}\biggr]\;,\;\;
\end{equation}
where $\Omega_{M,0}=\Omega_{DM,0}+\Omega_{b,0}$, and 
\begin{equation}\label{g1g2}
g_{1}=\frac{0.0783(\Omega_{b,0}h^{2})^{-0.238}}{1+39.5(\Omega_{b,0}h^{2})^{0.763}}\;,\hspace{0.3cm}g_{2}=\frac{0.560}{1+21.1(\Omega_{b,0}h^{2})^{1.81}}.
\end{equation}
An angular scale $l_{A}$ for the sound horizon at decoupling epoch is defined as
\begin{equation}\label{Acoustic_scale}
l_{A}(\mathbf{X})\equiv(1+z_{*})\frac{\pi D_{A}(z_{*},\mathbf{X})}{r_{s}(z_{*},\mathbf{X})}\,,\hspace{1cm}
\end{equation}
where $r_{s}(z_{*},\mathbf{X})$ is the comoving sound horizon at $z_{*}$, and is given by Eq. (\ref{rs}).
From \cite{Planck2015,Neveu2016}, the $\chi^{2}$ is
\begin{equation}\label{X2CMB}
\chi_{\bf CMB}^{2}(\mathbf{X})=\left({\Delta x}_{i}\right)^{t}\left(C^{-1}_{\bf CMB}\right)_{ij}\left({\Delta x}_{j}\right)\,,
\end{equation}
where ${\Delta x}_{i}=x^{\mathrm{th}}_{i}(\mathbf{X})-x^{\mathrm{obs}}_{i}$ is a column vector 
\begin{equation}
x^{\mathrm{th}}_{i}(\mathbf{X})-x^{\mathrm{obs}}_{i}=\left(\begin{array}{cc}
 l_{A}(z_{*})-301.7870\\
 R(z_{*})-1.7492\\
 \;\;z_{*}-1089.990\\
\end{array}\right)\;,
\end{equation}
``t'' denotes its transpose and $(C^{-1}_{\bf CMB})_{ij}$ is the inverse covariance matrix \cite{Neveu2016}
given by
\begin{equation}\label{MatrixCMB}
C^{-1}_{\bf CMB}\equiv\left(
\begin{array}{ccc}
+162.48&-1529.4&+2.0688\\
-1529.4&+207232&-2866.8\\
+2.0688&-2866.8&+53.572\\
\end{array}\right)\;.
\end{equation}
The errors for the CMB data are contained in $C^{-1}_{\bf CMB}$.
\begingroup
\squeezetable
\begin{table}[!htb]
\begin{tabular}{| c  | c  | c  | c  | c  | c  | c  | c | }
\hline
 $\mathrm{z}$   & $\mathrm{\bar{H}(z)}$ &  $1\sigma$& Refs. & $\mathrm{z}$   & $\mathrm{\bar{H}(z)}$ &  $1\sigma$& Refs. \\
\hline
$0.070$&  $69.0$&  $\pm19.6$&\cite{Zhang2014}    & $0.570$&  $96.8$&  $\pm3.40$&\cite{Anderson2014a}\\
$0.090$&  $69.0$&  $\pm12.0$&\cite{Simon2005}    & $0.593$& $104.0$&  $\pm13.0$&\cite{Moresco2012}\\
$0.120$&  $68.6$&  $\pm26.2$&\cite{Zhang2014}    & $0.600$&  $87.9$&  $\pm6.1$ &\cite{Blake2012}\\
$0.170$&  $83.0$&  $\pm8.0$&\cite{Simon2005}     & $0.680$&  $92.0$&  $\pm8.0$ &\cite{Moresco2012}\\
$0.179$&  $75.0$&  $\pm4.0$&\cite{Moresco2012}   & $0.730$&  $97.3$&  $\pm7.0$ &\cite{Blake2012}\\
$0.199$&  $75.0$&  $\pm5.0$&\cite{Moresco2012}   & $0.781$& $105.0$&  $\pm12.0$&\cite{Moresco2012}\\
$0.200$&  $72.9$&  $\pm29.6$&\cite{Zhang2014}    & $0.875$& $125.0$&  $\pm17.0$&\cite{Moresco2012}\\
$0.240$&  $79.69$& $\pm2.99$&\cite{Gastanaga2009}& $0.880$&  $90.0$&  $\pm40.0$&\cite{Stern2010}\\
$0.270$&  $77.0$&  $\pm14.0$&\cite{Simon2005}    & $0.900$& $117.0$&  $\pm23.0$&\cite{Simon2005}\\
$0.280$&  $88.8$&  $\pm36.6$&\cite{Zhang2014}    & $1.037$& $154.0$&  $\pm20.0$&\cite{Gastanaga2009}\\
$0.300$&  $81.7$&  $\pm6.22$&\cite{Oka2014}      & $1.300$& $168.0$&  $\pm17.0$&\cite{Simon2005}\\
$0.340$&  $83.8$&  $\pm3.66$&\cite{Gastanaga2009}& $1.363$& $160.0$&  $\pm33.6$&\cite{Moresco2015}\\
$0.350$&  $82.7$&  $\pm9.1$& \cite{Chuang2013a}  & $1.430$& $177.0$&  $\pm18.0$&\cite{Simon2005}\\
$0.352$&  $83.0$&  $\pm14.0$&\cite{Moresco2012}  & $1.530$& $140.0$&  $\pm14.0$&\cite{Simon2005}\\
$0.400$&  $95.0$&  $\pm17.0$&\cite{Simon2005}    & $1.750$& $202.0$&  $\pm40.0$&\cite{Simon2005}\\
$0.430$&  $86.45$& $\pm3.97$&\cite{Gastanaga2009}& $1.965$& $186.5$&  $\pm50.4$&\cite{Moresco2015}\\
$0.440$&  $82.6$&  $\pm7.8$&\cite{Blake2012}     & $2.300$& $224.0$&  $\pm8.6$ &\cite{Busca2013}\\
$0.480$&  $97.0$&  $\pm62.0$&\cite{Stern2010}    & $2.340$& $222.0$&  $\pm8.5$ &\cite{Debulac2015}\\
$0.570$&  $87.6$&  $\pm7.80$&\cite{Chuang2013b}  & $2.360$& $226.0$&  $\pm9.3$ &\cite{FontRibera2014}\\
\hline\end{tabular}
\caption{Shows the observational $\mathrm{\bar{H}(z)}$ data \cite{Chuang2013a,Chuang2013b,Anderson2014a,Debulac2015,FontRibera2014,Zhang2014,Simon2005,Moresco2012,Gastanaga2009,
Oka2014,Blake2012,Stern2010,Moresco2015,Busca2013}}\label{tableOHD} 
\end{table}
\endgroup
\begingroup
\squeezetable
\begin{table}[!htb]
\begin{tabular}{|c|@{\extracolsep{0mm}\ }c@{ }|}
\hline
Parameters&Constant Priors\\[0.2mm]
\hline
${\lambda}_{0}$&$[-1.5\times10^{+2},+1.5\times10^{+2}]$\\[0.2mm]
${\lambda}_{1}$&$[-1.5\times10^{+2},+1.5\times10^{+2}]$\\[0.2mm]
${\lambda}_{2}$&$[-1.5\times10^{+1},+1.5\times10^{+1}]$\\[0.2mm]
$\omega_{0}$&$[-2.0,-0.3]$\\[0.2mm]
$\omega_{1}$&$[-1.0,+1.0]$\\[0.2mm]
$\omega_{2}$&$[-2.0,+0.1]$\\[0.2mm]
$\Omega_{DM,0}$&$[0,0.7]$\\[0.2mm]
$\mathrm{H_{0}}(\mathrm{kms^{-1}{Mpc}^{-1}})$&$[20,120]$\\[0.2mm]
$\alpha$&$[-0.2,+0.5]$\\[0.2mm]
$\beta$&$[+2.1,+3.8]$\\[0.2mm]
$M$&$[-20,-17]$\\[0.2mm]
$dM$&$[-1.0,+1.0]$\\[0.2mm]
$\gamma_{0}$&$[+0.2,+1.2]$\\[0.2mm]
$\sigma_{80}$&$[0,+1.65]$\\[0.2mm]
\hline
\end{tabular} 
\caption{Shows the priors on the parameter space.\hspace{2.7cm}}\label{Priors}
\end{table}
\endgroup
\subsection{Observational Hubble data ($\mathrm{\bar{H}(z)}$)}\label{OHD}
Recently G. S. Sharov \cite{Sharov2015} compiled a list of $38$ independent measurements of the Hubble parameter at diffe-rent redshitfs, and used these measurements to 
constrain different cosmological models (see Table $\rm{III}$ in \cite{Sharov2015}). The $\chi^2_{\mathrm{H}}$ function for this data set is
\begin{equation}\label{X2OHD}
\chi^2_{\,\bf{\bm{H}}}(\mathbf{X})\equiv\sum_{i=1}^{38}\frac{\left[\mathrm{H}^{\rm {th}}(\mathbf{X},\mathrm{z_{i}},)-\mathrm{H}^{\mathrm{obs}}(\mathrm{z_{i}})\right]^2}{\sigma^2(\mathrm{z_{i}})}\;\;,
\end{equation}
where $\mathrm{H}^{{\rm th}}$ denotes the theoretical value for the Hubble parameter, $\mathrm{H}^{\mathrm{obs}}$ represents the observed value, 
$\sigma(\mathrm{z_{i}})$ is the standard deviation measurement uncertainty. This test has been used to constrain some models in \cite{Sharov2015}.\\
In order, to find  the best fit model-parameters, we perform a joint analysis using all the data, then we minimize 
\begin{equation}\label{TotalChi}
{{{\rm {\bf{\chi}}^{2}}}}={{{\rm{\bf{\chi}}^{2}}}}_{\bf JLA}+{{{\rm {\bf {\chi}}^{2}}}}_{\bf{RSD}}+{{{\rm {\bf {\chi}}^{2}}}}_{\bf BAO}+{{{\rm {\bf {\chi}}^{2}}}}_{\bf CMB}+
{{{\rm {\bf {\chi}}^{2}}}}_{\bf H}\;\;.
\end{equation}
Thus, the total probability density function ${{\rm {\bf pdf}}}$, is 
\begin{equation}\label{TotalexpChi}
{{\rm {\bf pdf}}}(\mathbf{X})=\rm{A}{{\rm e}}^{-{{\chi}}^{2}/2}\,.
\end{equation}
where $\rm A$ is a integration constant.
\subsection{Constant Priors}\label{Values}
In this work, we have assumed that baryonic matter and radiation are not coupled to DE or DM, which are separately conserved \cite{Koyama2009-Brax2010}. 
In this regard, we believe that the intensity of the interaction, ${\rm I}_{\rm Q}$, is not affected by the values of $\Omega_{b,0}$ and $\Omega_{r,0}$, respectively. 
Using these assumptions we can construct a ${{\rm {\bf pdf}}}$ function for our models. The priors considered here are given in Table \ref{Priors}.
\begin{figure*}[!htb]
 \includegraphics[width=9cm,height=4cm]{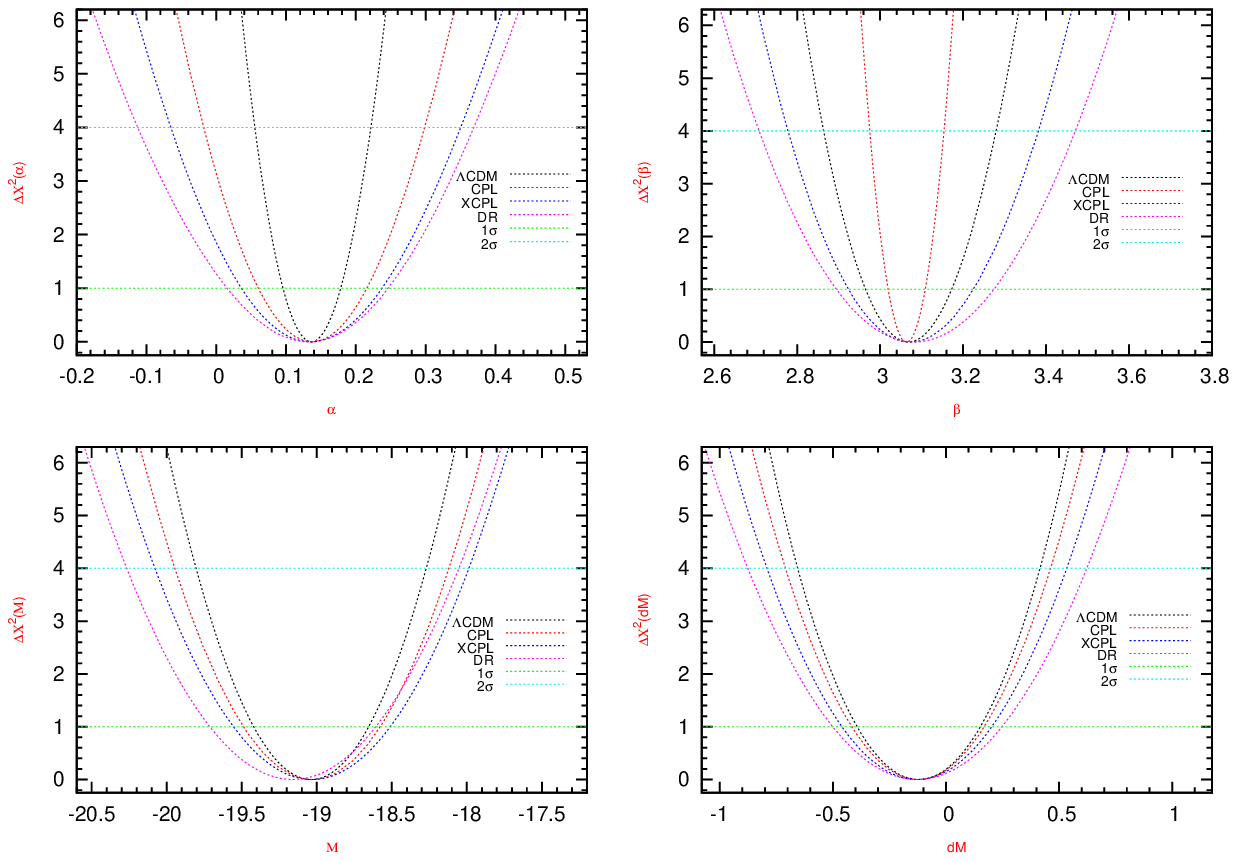}\includegraphics[width=9cm,height=4cm]{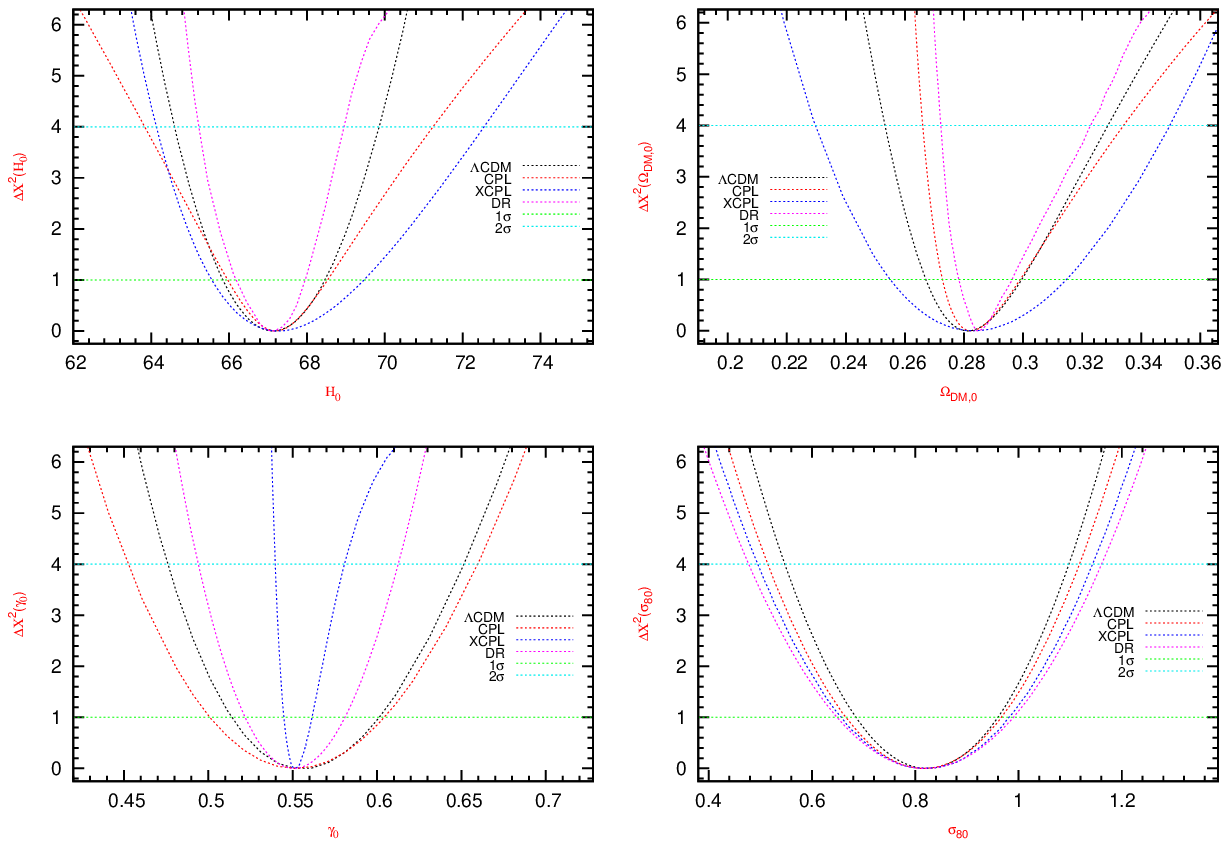}\\
 \includegraphics[width=9cm,height=4cm]{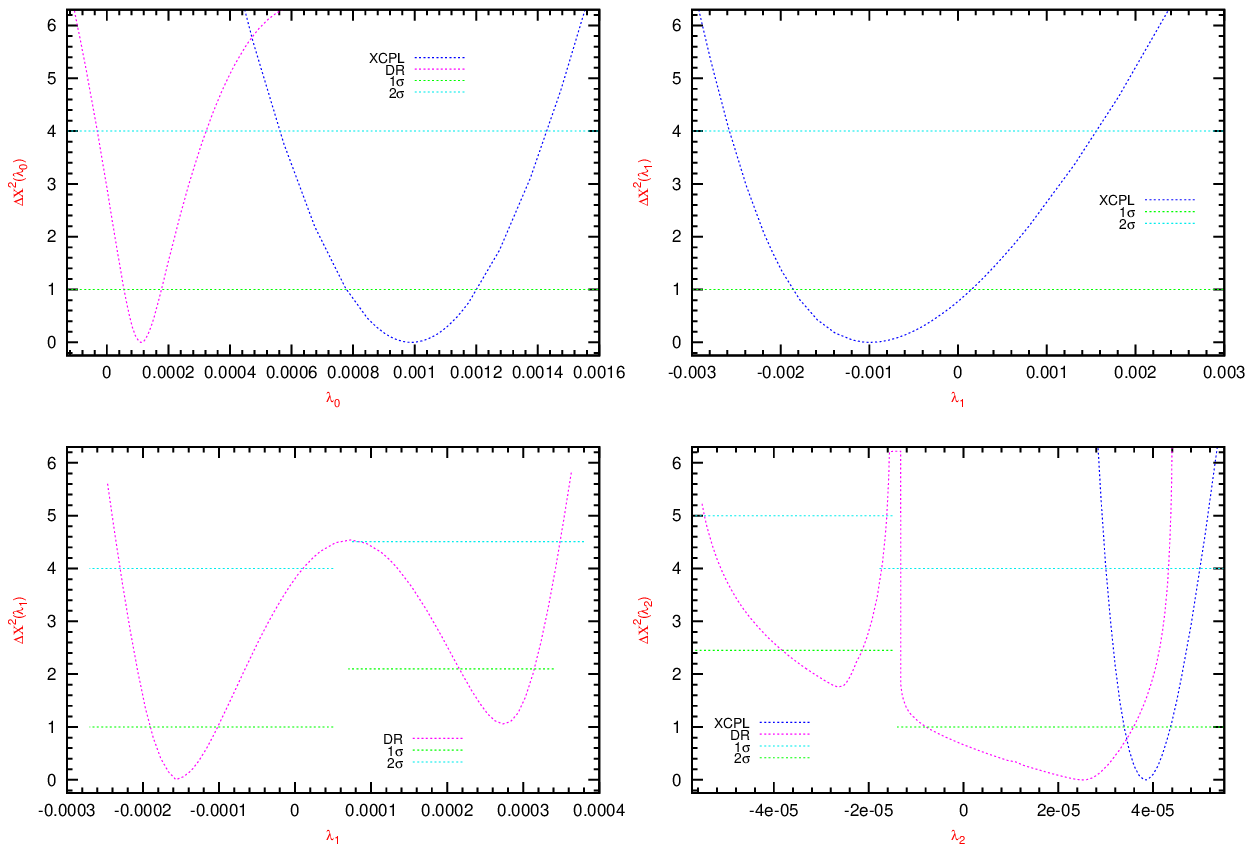}\includegraphics[width=9cm,height=4cm]{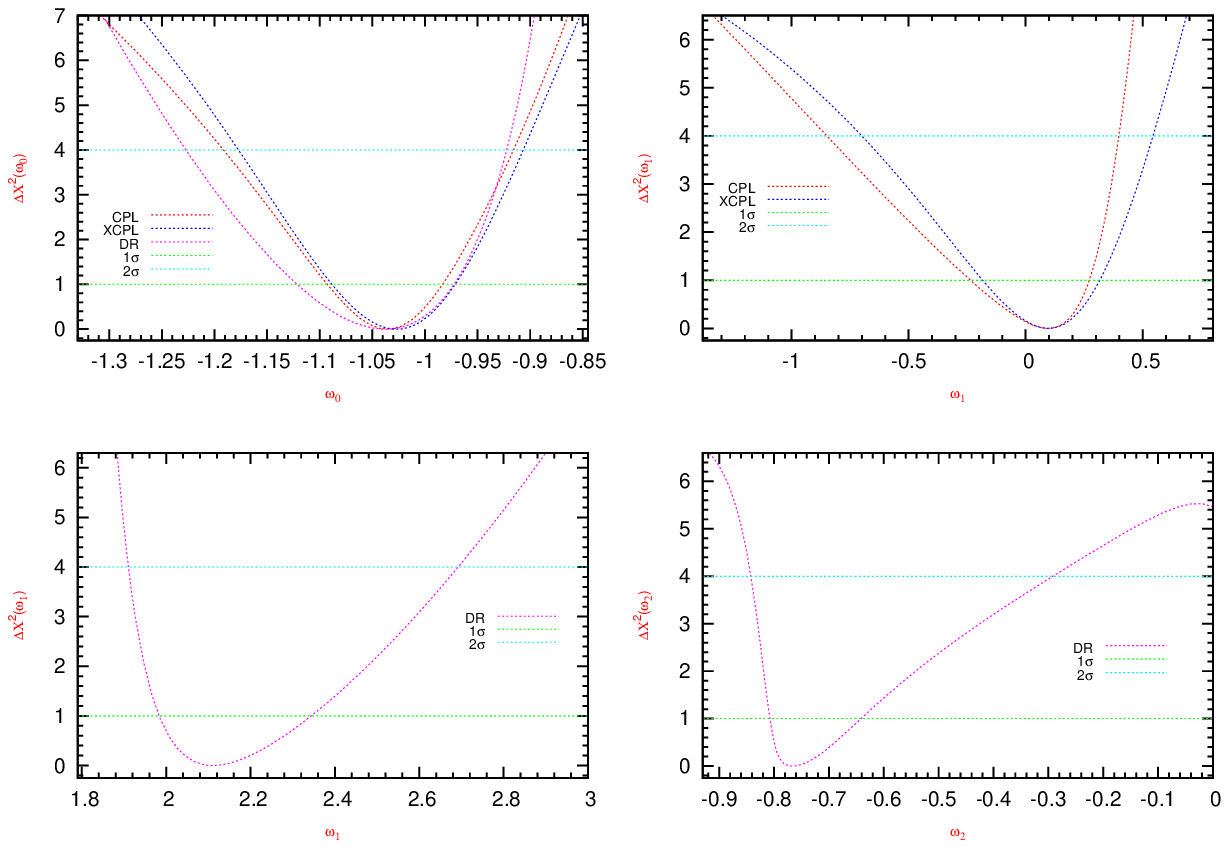}\\
 \caption{(color online) \label{Contours} Displays the one-dimension probability contours for all the parameters worked and their constraints at $1\sigma$ and $2\sigma$, respectively. 
 Here, we consider that $\Delta{\chi^{2}}=\chi^{2}-{\chi^{2}_{min}}$\,.}
\end{figure*} 
\begingroup
\squeezetable
\begin{table*}[!hbtp]
\centering
\caption{Shows the best fitting cosmological parameters for each model and their constraints at $1\sigma$ and $2\sigma$ obtained from an analysis of 
Union JLA+RSD+BAO+CMB+H data sets.}\label{Bestfits}
\begin{tabular}{| c | c | c | c | c | c |}
\hline
 Parameters&$\Lambda$CDM&CPL&XCPL&DR(1)&DR(2)\\
 \hline\hline 
${\lambda}_{0}\times10^{+4}$&$N/A$&$N/A$&${+9.855}^{+2.1791+4.4462}_{-2.0866-4.2642}$&${+1.120}^{+0.6525+2.0844}_{-0.6004-4.2761}$&${+1.12}^{+0.6525+2.0844}_{-0.6004-4.2761}$\\[0.4mm]
${\lambda}_{1}\times10^{+4}$&$N/A$&$N/A$&${-9.870}^{+11.4244+25.5981}_{-8.7497-15.8883}$&${+2.733}^{+0.4106+0.7486}_{-0.5670-1.8947}$&${+2.733}^{+0.4106+0.7486}_{-0.5670-1.8947}$\\[0.4mm]
${\lambda}_{2}\times10^{+5}$&$N/A$&$N/A$&${+3.850}^{+0.5279+1.1475}_{-0.4597-0.8276}$&${+2.539}^{+1.0254+1.7806}_{-10.2680-3.8349}$&${-2.649}^{+0.5207+1.0435}_{-1.1834-2.8293}$\\[0.6mm]
$\omega_{0}$&$-1.0$&${-1.0323}^{+0.0489+0.1165}_{-0.0605-0.1586}$&${-1.0271}^{+0.0563+0.1208}_{-0.0610-0.1497}$&${-1.0364}^{+0.0644+0.1140}_{-0.0853-0.1908}$&${-1.0364}^{+0.0644+0.1140}_{-0.0853-0.1908}$\\[0.4mm]
$\omega_{1}$&$N/A$&${+0.0952}^{+0.1757+0.3024}_{-0.3267-0.9446}$&${+0.0950}^{+0.2218+0.4488}_{-0.2827-0.7960}$&${+2.1064}^{+0.2363+0.5842}_{-0.1213-0.1964}$&${+2.1064}^{+0.2363+0.5842}_{-0.1213-0.1964}$\\[0.4mm]
$\omega_{2}$&$N/A$&$N/A$&$N/A$&${-0.7698}^{+0.1276+0.4797}_{-0.0364-0.0717}$&${-0.7698}^{+0.1276+0.4797}_{-0.0364-0.0717}$\\[0.4mm]
$\Omega_{DM,0}$&${+0.2810}^{+0.0185+0.0476}_{-0.0138-0.0279}$&${+0.2814}^{+0.0176+0.0528}_{-0.0089-0.0154}$&${+0.2840}^{+0.0308+0.0659}_{-0.0290-0.0542}$&${+0.2844}^{+0.0121+0.0385}_{-0.0061-0.0124}$&${+0.2844}^{+0.0121+0.0385}_{-0.0061-0.0124}$\\[0.4mm]
$\mathrm{H_{0}}(\mathrm{\frac{km}{s.Mpc}})$&${+67.170}^{+1.274+2.694}_{-1.3079-2.5666}$&${+67.19}^{+1.3508+4.0403}_{-1.2203-3.3504}$&${+67.20}^{+2.2767+5.3535}_{-1.6607-3.0699}$&${+67.1490}^{+0.8216+1.8006}_{-0.9642-1.9324}$&${+67.1490}^{+0.8216+1.8006}_{-0.9642-1.9324}$\\[0.4mm]
$\alpha$&${+0.1360}^{+0.0419+0.0855}_{-0.0410-0.0814}$&${+0.1370}^{+0.0787+0.1621}_{-0.0758-0.1542}$&${+0.1350}^{+0.1017+0.2148}_{-0.0992-0.1993}$&${+0.1360}^{+0.1108+0.2341}_{-0.1198-0.2482}$&${+0.1360}^{+0.1108+0.2341}_{-0.1198-0.2482}$\\[0.4mm]
$\beta$&${+3.068}^{+0.1033+0.2129}_{-0.1026-0.2035}$&${+3.065}^{+0.0434+0.0903}_{-0.0465-0.0897}$&${+3.078}^{+0.1462+0.2953}_{-0.1556-0.3010}$&${+3.0780}^{+0.1968+0.3939}_{-0.1839-0.370}$&${+3.0780}^{+0.1968+0.3939}_{-0.1839-0.3700}$\\[0.4mm]
$M$&${-19.0340}^{+0.3849+0.7605}_{-0.3907-0.7690}$&${-19.030}^{+0.4560+0.9122}_{-0.4591-0.9179}$&${-19.0310}^{+0.5241+1.0447}_{-0.5270-1.6457}$&${-19.1650}^{+0.5561+1.1116}_{-0.5522-1.0996}$&${-19.1650}^{+0.5561+1.1116}_{-0.5522-1.0996}$\\[0.4mm]
$dM$&${-0.120}^{+0.0299+0.2983}_{-0.2718-0.5360}$&${-0.121}^{+0.2907+0.5838}_{-0.2975-0.5906}$&${-0.125}^{+0.3326+0.6632}_{-0.3365-0.6633}$&${-0.125}^{+0.3715+0.7488}_{-0.3832-0.7539}$&${-0.125}^{+0.3715+0.7488}_{-0.3832-0.7539}$\\[0.4mm]
$\gamma_{0}$&${+0.5511}^{+0.0506+0.1010}_{-0.0375-0.0753}$&${+0.5510}^{+0.0529+0.1088}_{-0.0506-0.0985}$&${+0.5510}^{+0.0110+0.0298}_{-0.0010-0.0119}$&${+0.5511}^{+0.0302+0.0615}_{-0.0291-0.0571}$&${+0.5511}^{+0.0302+0.0615}_{-0.0291-0.0571}$\\[0.4mm]
$\sigma_{80}$&${+0.8180}^{+0.1400+0.2794}_{-0.1340-0.2718}$&${+0.8190}^{+0.1471+0.2987}_{-0.1504-0.3036}$&${+0.8180}^{+0.1643+0.3257}_{-0.1624-0.3213}$&${+0.8190}^{+0.1706+0.3425}_{-0.1690-0.3417}$&${+0.8190}^{+0.1706+0.3425}_{-0.1690-0.3417}$\\[0.4mm]
\hline\hline${\chi}^{2}_{min}$&$737.8581$&$736.5633$&$723.8474$&$712.3048$&$723.7758$\\[0.4mm]
\hline
\end{tabular}
\end{table*}
\endgroup
\begin{figure*}[!htb]
 \centering \includegraphics[width=18.5cm,height=8.5cm]{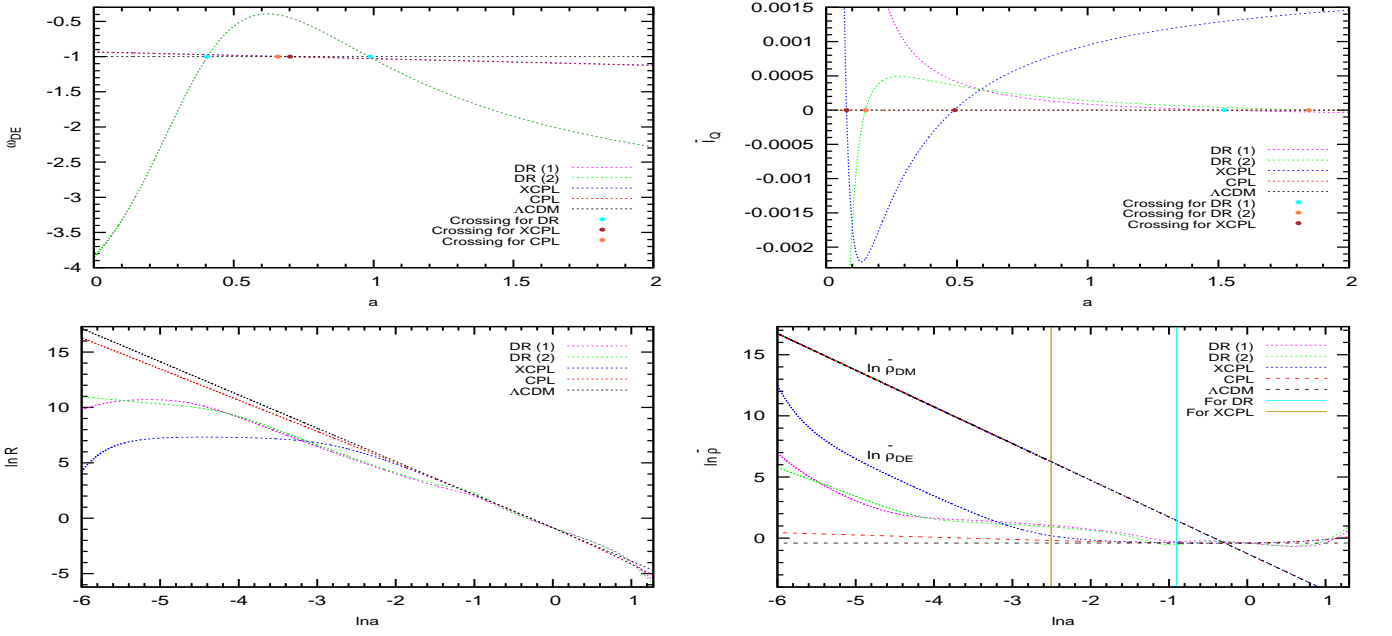}
 \caption{(color online) \label{Reconstructions} Shows the background evolution of $\mathrm{\omega_{DE}}$ and ${\bar{\rm I}}_{\rm Q}$, $\mathrm{R}$ and $\mathrm{\bar{\rho}}$ along $\mathrm{a}$ for the coupled and 
 uncoupled models. From the left below panel, we note that $\mathrm{R}$ is always positive when both $\bar{{\rm I}}_{\rm Q}$ and $\mathrm{\omega_{DE}}$ are time-varying, and remains finite 
 when $\mathrm{a}\rightarrow \infty$. As is apparent, $\bar{Q}$ seems to alleviate the coincidence problem for ln$\mathrm{a}\leq 0$. In addition, in the right below panel, the vertical lines 
 indicate the moment when $|3\bar{\mathcal{H}}(1+\mathrm{\omega_{DE}})\mathrm{{\bar\rho}_{DE}}|$ and $|\mathrm{{\bar\rho}_{DM}}{\bar{\rm I}}_{\rm Q}|$ are equal, see Eq. (\ref{EDE}). Here, to the left 
 of theses lines, $\bar{Q}$ affects the background evolution of $\mathrm{\bar{\rho}_{DE}}$. By contrast, the situation is opposite to the right of these lines.} 
\end{figure*}
\begin{figure*}[!htb]
\centering\includegraphics[width=18.5cm,height=8.5cm]{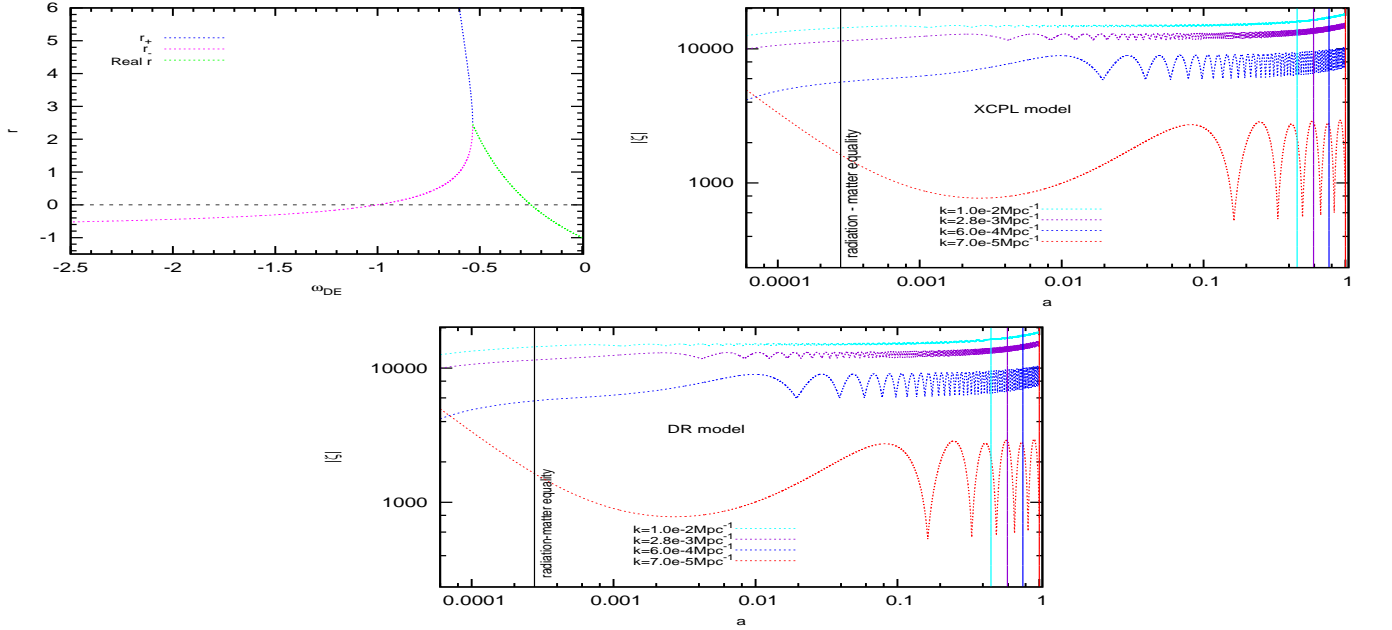}
 \caption{(color online) \label{Effects1} 
 The left above panel displays the evolution of the index r given by Eq. (\ref{indices}) as function of $\mathrm{\omega_{DE}}$. These three graphs meet at the point 
 where $\sqrt{10\mathrm{\omega_{DE}}^{2}-4\mathrm{\omega_{DE}}-5}=0$ and their real parts are identical (green color line). The right above and below panels show the evolution of the gauge-
 invariant curvature perturbation $|\mathrm{\zeta}|$ in the coupled models for four different scales as function of $\mathrm{a}$. The vertical lines indicate the moment when each mode enters 
 the horizon ($\mathrm{k}\tau\sim{1}$) and the moment when the radiation and matter eras are equal (black line). The largest scale ($\mathrm{k}=\mathrm{7.0\times 10^{-5}Mpc^{-1}}$) stays at large-
 scales all the way-up to today. The intermediate scales ($\mathrm{k}=\mathrm{6.0\times 10^{-4}Mpc^{-1}}$, $\mathrm{k}=\mathrm{2.8\times 10^{-3}Mpc^{-1}}$ and $\mathrm{k}=\mathrm{1.0\times 10^{-2}Mpc^{-1}}$) enters the horizon 
 during matter dominated era. These curves were obtained from our codes in $c^{++}$ language, with the initial amplitude of $\psi=10^{-25}$.}
  \end{figure*}
 \begin{figure*}[!htb]
 \centering \includegraphics[width=18.5cm,height=8.5cm]{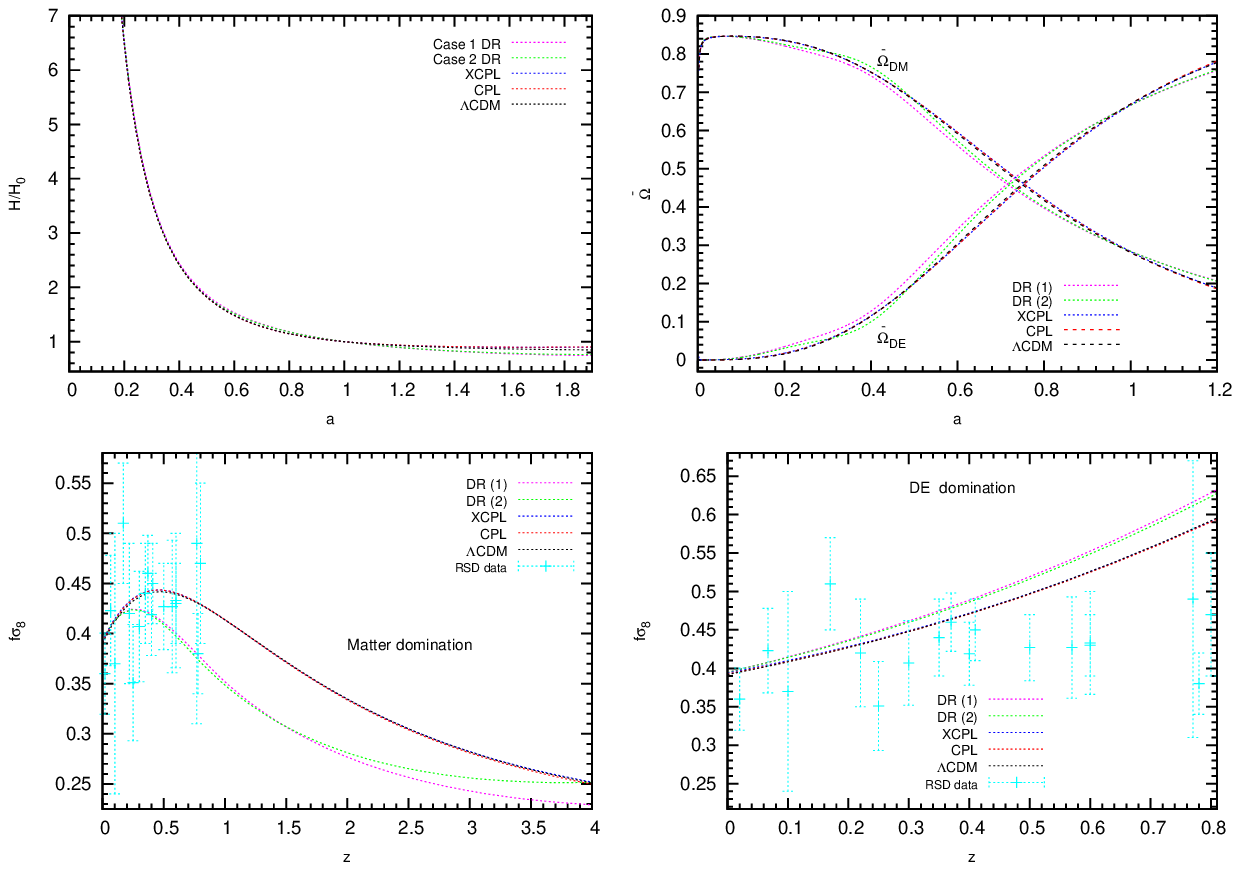}
 \caption{(color online) \label{Effects2} 
 The left above panel shows the evolution of the Hubble parameter normalized to $\mathrm{H_{0}}$ along $\mathrm{a}$. Here, a slight deviation for $\mathrm{H/H_{0}}$ in the DR model respect to other 
 models is shown for $\mathrm{a}\geq 1$. The right above panel depicts the evolution of $\mathrm{\bar{\Omega}_{DM}}$ and $\mathrm{\bar{\Omega}_{DE}}$ as function of a. Here, the XCPL, CPL and 
 $\Lambda$CDM scenarios have slightly higher $\mathrm{{\bar{\Omega}}_{DM}}$ (slower $\mathrm{{\bar{\Omega}}_{DE}}$) from the past to today compared to that found in the DR model. By contrast, 
 the situation is opposite in the future. From the left below panel, we note in the DR model a suppression on the amplitude of $\mathrm{f\sigma_{8}}$ in comparison with other models 
 and during the matter era. In the right below panel, such behaviour for $\mathrm{f\sigma_{8}}$ is different inside the DE domination epoch.}
\end{figure*}
\section{Model Selection Statistics}
In this section, to know whether a model is favored by data, we will use selection methods statistics such as:\\ 
a) $\chi^{2}/dof$ ($dof$: degrees of freedom),\\ 
b) Goodness of Fit ($GoF$), simply gives the probability of obtaining, by chance, a data set that is a worse fit to the model than the actual data, assuming that the model is correct. It is defined as:
\begin{equation}\label{GoF} 
GoF \equiv \frac{\Gamma\left(dof/2,\chi^{2}/2\right)}{\Gamma\left(dof/2\right)}\;,
\end{equation}
where $\Gamma$ is the incomplete gamma function.\\
c) For a model with $p$ free parameters, with a best-fit $\chi^{2}_{min}$, and for a finite number of data points $M$ used in the fit, the $AIC$ and $BIC$ criteria are 
\begin{eqnarray}\label{AIC} 
AIC \equiv \chi^{2}_{min} + 2p\;,\\
BIC \equiv \chi^{2}_{min} + p\ln\,M\,.
\end{eqnarray}
The model with the the smaller value of $AIC$ (or $BIC$) is usually considered to be the best (the preferred model). The absolute value of the criterion for a single model 
has no meaning and only the relative values between diffe-rent models are interesting. 
This difference compares the model $i$ with the model $j$ through $\Delta AIC_{ij}=AIC_{i}-AIC_{j}=\Delta\chi^{2}_{min}+2\Delta p$ 
(or $\Delta BIC_{ij}=BIC_{i}-BIC_{j}=\Delta\chi^{2}_{min}+2\Delta p\ln\,M$). The results for the $AIC$ criterion are: $0\leq \Delta AIC_{ij}\leq 2$, the model $i$ has 
a ``strong evidence in favour'' from the data, for $4\leq \Delta AIC_{ij}\leq 7$, the model $i$ has a ``little evidence in favour'', and with 
$\Delta AIC_{ij}\geq 10$. The model $i$ is practically irrelevant (``no evidence in favour''). Similarly, the $\Delta BIC_{ij}$, can be understood as ``evidence against'' 
the model $i$ compared to the model $j$. For $0\leq \Delta BIC_{ij}\leq 2$, the model $i$ has not ``enough evidence against'' from the data, for 
$2\leq \Delta BIC_{ij}\leq 6$, the model $i$ has an ``evidence against'', and for $6\leq \Delta BIC_{ij}\leq 10$, the model $i$ has a ``strong evidence against'' from the 
observational data.
\begingroup
\squeezetable
\begin{table*}[!hbtp]
 \begin{tabular}{>{\centering\arraybackslash}m{10.5cm} >{\arraybackslash}m{10.5cm}}
{
\begin{tabular}{| c | c | c | c | c | c | c | c | c | c |}
\hline
Models&$p$&$M$&$\chi^{2}_{min}$&$\chi^{2}_{min}/dof$&$GoF$&$AIC$&$BIC$&$|\Delta AIC|$&$|\Delta BIC|$\\
\hline\hline
$\Lambda$CDM&$9$&$826$&$737.8581$&$0.9031$&$0.9777$&$755.8581$&$798.3075$&$0.0$&$0.0$\\[0.35mm]
CPL&$10$&$826$&$736.5633$&$0.9027$&$0.9782$&$756.5633$&$803.7293$&$0.7052$&$5.4218$\\[0.35mm]
XCPL&$13$&$826$&$723.8474$&$0.8903$&$0.9887$&$749.8474$&$811.1632$&$6.0107$&$12.8557$\\[0.35mm]
DR1&$14$&$826$&$712.3048$&$0.8772$&$0.9949$&$740.3048$&$806.3371$&$15.5534$&$8.0296$\\[0.35mm]
DR2&$14$&$826$&$723.7758$&$0.8914$&$0.9880$&$751.7758$&$817.8081$&$4.0824$&$19.5006$\\[0.35mm]
\hline
\end{tabular}
\caption{The $\Lambda$CDM model has the lower $AIC$ and $BIC$ values.}\label{Criteria1}
} & 
{
\begin{tabular}{| c | c | c | c | c | c | c | c |}
\hline
Models&$p$&$\chi^{2}_{min}$&$AIC$&$BIC$&$|\Delta AIC|$&$|\Delta BIC|$\\
\hline\hline
DR1&$14$&$712.3048$&$740.3048$&$806.3371$&$0.0$&$0.0$\\[0.35mm]
XCPL&$13$&$723.8474$&$749.8474$&$811.1632$&$9.5427$&$4.8261$\\[0.35mm]
DR2&$14$&$723.7758$&$751.7758$&$817.8081$&$11.4710$&$11.4710$ \\[0.35mm]
\hline
\end{tabular}
\caption{The DR1 model is the underlying.\hspace{4cm}}\label{Criteria2}
}
\end{tabular}
\end{table*}
\endgroup
\section{Results}
We have built all the codes required to calculate numerically the theoretical evolutions of $\mathrm{R}$, $\mathrm{{\bar{\rho}}_{A}}$, $|\delta_{A}|$, $\mathrm{f\sigma_{8}}$ and $|\mathrm{\zeta}|$ with 
$A=DM, DE$, in our models and via a Markov Chain Monte Carlo (MCMC) analysis, we have performed a global fitting on each one of them (listed in Table \ref{Bestfits}), 
by using current observational data. Table \ref{Priors} describes the priors used in this work. For each one of the models, the one-dimension probability contours, the 
best-fit parameters and their errors (at $1\sigma$ and $2\sigma$) are shown in Fig. \ref{Contours}.\\
In the following Figs. the constraints at $1\sigma$ and $2\sigma$ on $\mathrm{\omega_{DE}}$, ${\bar{\rm I}}_{\rm Q}$, $\mathrm{R}$, $\mathrm{{\bar{\rho}}_{DM}}$, $\mathrm{{\bar{\rho}}_{DE}}$, 
$\mathrm{f\sigma_{8}}$, and $\mathrm{\zeta}$ have been omitted to obtain a better visualization of the results.
Due to the two minimums obtained in the DR model (see Table \ref{Bestfits}), two different cases ($1$ and $2$) to reconstruct ${\rm I}_{\rm Q}$ are worked here.
Furthermore, in the left above panel of Fig. \ref{Reconstructions} the universe evolves from the quintessence regime $\mathrm{\omega_{DE}} > -1$ to the phantom regime 
$\mathrm{\omega_{DE}} < -1$ or vice versa, and in particular, crosses the phantom divide line $\mathrm{\omega_{DE}}=-1$ \cite{Nesseris2007}. The DR model has two crossing points 
in $\mathrm{a}=0.4043$ and $\mathrm{a}=0.9894$, respectively. Instead, the XCPL model shows only one in $\mathrm{a}=0.7015$. Likewise, the CPL model also depicts one in $\mathrm{a}=0.6581$. From this 
panel, we stress that there is a significant difference for the evolution of $\mathrm{\omega_{DE}}$ in the XCPL and DR models. Besides, the behaviour of $\mathrm{\omega_{DE}}$ in the DR model 
is opposite with that showed in \cite{Lu2009-Neveu2013}.\\
Let us now see the right above panel of Fig. \ref{Reconstructions}. Here, within the coupled models have considered that ${\rm I}_{+}$ denotes an energy transfer from DE 
to DM and ${\rm I}_{-}$ denotes an energy transfer from DM to DE and have found a change from ${\rm I}_{+}$ to ${\rm I}_{-}$ and vice versa. This change of sign is 
linked to ${\bar{\rm I}}_{\rm Q}=0$. The XCPL model has two crossing points in $\mathrm{a}=0.0784$ and $\mathrm{a}=0.4922$; instead, the DR model shows three crossing points in 
$\mathrm{a}=1.5238$ (DR 1), $\mathrm{a}=0.1512$ and $\mathrm{a}=1.8462$ (DR 2), respectively. These points were already predicted by Eq. (\ref{ZcrossIq}). From Table \ref{Bestfits} we 
note that DR 2 is in disagreement with the result obtained in Eq. (\ref{dIqdznull6}).\\ 
The left below panel of Fig. \ref{Reconstructions} shows the background evolution of ln$\mathrm{R}$ which exhibits a scaling behaviour for ln$\mathrm{R}$, keeping constant at early times 
but not at present times of the universe. These results significantly alleviate the coincidence problem (see Eq. (\ref{alleviate}), but they do not solve it in full. 
From the right below panel of the Fig. \ref{Reconstructions}, we see that at ln$\mathrm{a}<0$ the coupling affects violently the background evolution of $\mathrm{{\bar{\rho}}_{DE}}$ in 
the DR and XCPL models. By contrast, the situation is opposite at ln$\mathrm{a}>0$. Furthermore, the graphs for $\mathrm{{\bar{\rho}}_{DM}}$ are essentially overlap during their evolution.\\
The left above panel of Fig.\,\ref{Effects1} shows that when $\mathrm{\omega_{DE}}<-1$ or $-0.25<\mathrm{\omega_{DE}}<0$, and both r$_{-}$ or Real r are real negative, then, the DE 
perturbation does not diverge. Nevertheless, when $-1<\mathrm{\omega_{DE}}<-0.25$, and r$_{+}$, r$_{-}$ and Real r are real positive, then, the DE perturbation diverges violently 
(an instability when $\mathrm{\omega_{DE}}\neq-1$). These three graphs meet at the point where $\sqrt{10\mathrm{\omega_{DE}}^{2}-4\mathrm{\omega_{DE}}-5}=0$, and their real parts are identical. 
This analysis is independent of ${\bar{\rm I}}_{\rm Q}$.\\
We see from the right above and below panels of Fig. \ref{Effects1} that when $\mathrm{\omega_{DE}}$ and ${\bar{\rm I}}_{\rm Q}$ are time-varying and for four different values 
of $\mathrm{k}$, a violent growth of $|\mathrm{\zeta}|$ (so-called blow-up) in the coupled models during the radiation era. Here, $\bar{{\rm I}}_{\rm Q}$ is responsible 
of the instability why affects the background evolution. Moreover, when this background starts to behave as uncoupled, the blow-up of $|\mathrm{\zeta}|$ ends and it begins to 
oscillate with a determinated amplitude. Likewise, we have also verified that this instability does not depend on the sign of $\bar{{\rm I}}_{\rm Q}$. These results are in 
agreement with the analytical predictions given in left above panel of this Fig. We can also say that the results obtained in both cases of the DR model are identical 
and coincide with that obtained in the XCPL model. These results are very closed to those obtained in \cite{valiviita2008,Jackson2009,He2009,Xu2011}.\\
Within the matter era, we see that the amplitude of $\mathrm{f\sigma_{8}}$ in the DR model is suppressed relative to other models. Thus, in the XCPL model this behaviour is very 
closed with that predicted in the uncoupled models. These results are in agreement with the background evolution of $\mathrm{\bar{\Omega}_{DM}}$ in Fig. \ref{Effects2}. A lesser 
concentration $\mathrm{\bar{\Omega}_{DM}}$ leads to a lesser amplitude of $\mathrm{f\sigma_{8}}$. These effects are clearly determined by the background evolution of both $\bar{Q}$ and 
$\mathrm{\omega_{DE}}$ through the different background evolution of $\mathrm{\bar{\rho}_{DM}}$, $\mathrm{\bar{\rho}_{DE}}$, $\mathrm{\bar{H}}$ and $\delta_{DM}$, respectively. By contrast, the behaviour of 
$\mathrm{f\sigma_{8}}$ is opposite inside the DE era (see right below panel).\\
We now compare our results of $\mathrm{f\sigma_{8}}$ with those obtained by other researchers. In \cite{Alcaniz2013}, the authors provided a convenient analytic formula for 
$\mathrm{f\sigma_{8}}$, which was applied to different DE models. They used RSD data to place observational constraints. The results obtained by them on $\mathrm{f\sigma_{8}}$ are 
consistent at $1\sigma$ error with our result. Simi-larly, Yang and Xu in \cite{Yang2014}, studied a model composed by the cosmological constant, with a nonzero DM EoS 
parameter.\,Its result obtained on $\mathrm{f\sigma_{8}}$ is consistent with our result at $1\sigma$ error. Also, the authors in \cite{Mehrabi2015}, studied the impact of DE 
clustering on $\gamma$. They used two different EoS parameters, and found a best fit for $\mathrm{f\sigma_{8}}$, which at $1\sigma$ is acceptable with our result.\\
On the other hand, one can see from the values in Table \ref{Criteria1} that the $\Lambda$CDM model is strongly preferred. From here, $AIC$ method shows that the coupled models have ``little (no) 
evidence in favour''. Instead, for the $BIC$ selection method, it seems that the coupled models are not particularly favored over the $\Lambda$CDM model (because 
$\Delta BIC_{ij}\geq 6$). This last method tends to penalize the coupled models when are compared with the $\Lambda$CDM model because they have a larger number of parameters 
\cite{BIC_Criterion}. Moreover, if we compare only the XCPL and DR models and assume that the DR1 model is the underlying model. Then, the $AIC$ and $BIC$ criteria show that 
the DR1 model is the best (see Table \ref{Criteria2}).\\
\section{Conclusions}\label{SectionConclusions}             
Now we summarize our main results:\\
$\bullet$ An analysis combined of data was performed to break the degeneracy among the different parameters of our models, obtaining constraints more 
stringent on them.\\
$\bullet$ In the DR model, the reconstruction of $\mathrm{\omega_{DE}}$ has the property of avoiding divergences at $\mathrm{z}\rightarrow -1$ and its best-fit is consistent with 
the value predicted by the $\Lambda$CDM model at $1\sigma$ error.\\
$\bullet$ From the above results, we note that the coupling given by Eq. (\ref{Interaction}) fulfills the criteria of Sec. II provided for $\bar{Q}$ in 
\cite{Campo-Herrera2015}, and therefore, it is physically acceptable in the dark sector. \\
$\bullet$ In the coupled DE models the amplitudes of $\mathrm{{\bar{\rho}}_{DE}}$ are modified relative to those found in the uncoupled models. However, they are definitely 
positive. This implies that $\mathrm{\omega_{DE}}<0$ (see right below panel in Fig. \ref{Reconstructions}).\\
$\bullet$ For different values of $\mathrm{k}$ and with both time-varying $\mathrm{\omega_{DE}}$ and ${\bar{\rm I}}_{\rm Q}$, we note in the coupled models that the evolution of $|\mathrm{\zeta}|$ 
(see panels in Fig. \ref{Effects1}) follow similar behaviors but their amplitudes present a violent growth at early times (blow-up), and after
it begins to oscillate with a determinated amplitude. Moreover, this instability does not depend on the sign of $\bar{{\rm I}}_{\rm Q}$, the blow-up occurs even if 
${\bar{\rm I}}_{\rm Q}$ is very weak, when $-1<\mathrm{\omega_{DE}}<-0.25$. It has been verified numerically for us. The curvature perturbation have the possibility to be more 
stable when $\mathrm{\omega_{DE}}<-1$. However, this stability is absent for $-0.25<\mathrm{\omega_{DE}}<0$.\\
$\bullet$ A slight enhancement or suppression on the amplitudes of $\mathrm{\bar{\Omega}_{DM}}$, $\mathrm{\bar{\Omega}_{DE}}$ and $\mathrm{f\sigma_{8}}$ in the coupled models respect to the standard 
evolution show us the impact of $\bar{Q}$ and $\mathrm{\omega_{DE}}$ on them (see above and the left below panels in Fig. \ref{Effects2}).\\
$\bullet$ From the above and the right below panels in Fig. \ref{Effects2} we note that inside the DE era, the DM structure formation is slowed and stopped, 
due to the accelerating expansion of the universe and to the uniform creation of new DM via $\bar{Q}$. Besides, The amplitude of $\mathrm{f\sigma_{8}}$ shows a slight departure 
in the DR model from that found in the other models. Such feature is clearly different in the XCPL model.\\
$\bullet$ Finally, from the values of $AIC$ and $BIC$ presented in Table \ref{Criteria1}, we conclude that the XCPL and DR models are not preferred by the data \cite{Kurek2014}. 
Instead, the $\Lambda$CDM model is the best. But when the coupled models are only considered, we find that the DR1 model is favored by the data. This result is also verified    
by the $\chi^{2}/dof$ and $GoF$ methods, see Table \ref{Criteria2}.\\
\begin{acknowledgments}
The author is indebted to the Instituto de F\'{\i}sica y Matem\'aticas (UMSNH) for its hospitality and support.
\end{acknowledgments}


\begin{thebibliography}{78}
\begin{scriptsize}
\bibitem{Conley2011}
Conley A et al., \textit{Astrophys. J. Suppl.} \textbf{192} (2011) 1.
\bibitem{Jonsson2010}
J\"onsson, J., et al. 2010, \textit{Mon. Not. Roy. Astron. Soc.} \textbf{405} (2010) 535.
\bibitem{Betoule2014}
Betoule M et al., \textit{Astron. and Astrophys.} \textbf{568} (2014) A22.
\bibitem{Jackson1972}
J. C. Jackson, \textit{Mon. Not. Roy. Astron. Soc.} \textbf{156} (1972) 1P.
\bibitem{Kaiser1987}
Kaiser N., \textit{Mon. Not. Roy. Astron. Soc.} \textbf{227} (1987) 1.
\bibitem{Mehrabi2015}
A. Mehrabi, S. Basilakos, F. Pace, \textit{Mon. Not. Roy. Astron. Soc.} \textbf{452} (2015) 2930-2939.
\bibitem{Alcock1979}
Alcock C. and Paczynski B., \textit{Nature.} \textbf{281} (1979) 358. 
\bibitem{Seo2008}
H-J. Seo, E. R. Siegel, D. J. Eisenstein, and M. White, \textit{Astrophys. J.} \textbf{686} (2008) 13Y24.
\bibitem{Battye2015}
R. A. Battye, T. Charnock and A. Moss \textit{Phys. Rev.} \textbf{D 91} (2015) 103508.
\bibitem{Samushia2014}
L. Samushia, et al., \textit{Mon. Not. Roy. Astron. Soc.} \textbf{439} (2014) 3504.
\bibitem{Hudson2013}
Hudson M. J., Turnbull S. J., \textit{Astrophys. J.} \textbf{751} (2013) L30.
\bibitem{Beutler2012}
Beutler F., Blake C., Colless M., Jones D. H., Staveley-Smith L., et al., \textit{Mon. Not. Roy. Astron. Soc.} \textbf{423} (2012) 3430.
\bibitem{Feix2015}
Feix M., Nusser A., Branchini E., \textit{Phys. Rev. Lett.} \textbf{115} (2015) 011301.
\bibitem{Percival2004}
Percival W. J., et al., \textit{Mon. Not. Roy. Astron. Soc.} \textbf{353} (2004) 1201.
\bibitem{Song2009}
Y-S. Song and W. J. Percival, \textit{J. Cosmol. Astropart. Phys.} \textbf{10} (2009) 004.
\bibitem{Tegmark2006}
Tegmark M. et al., \textit{Phys. Rev.} \textbf{D 74} (2006) 123507.
\bibitem{Guzzo2008}
Guzzo L. et al., \textit{Nature.} \textbf{451} (2008) 541. 
\bibitem{Samushia2012}
Samushia L., Percival W. J., Raccanelli A., \textit{Mon. Not. Roy. Astron. Soc.} \textbf{420} (2012) 2102.
\bibitem{Blake2011}
Blake C. et al., \textit{Mon. Not. Roy. Astron. Soc.} \textbf{415} (2011) 2876; \textit{Mon. Not. Roy. Astron. Soc.} \textbf{418} (2011) 1725.
\bibitem{Tojeiro2012}
Tojeiro R., Percival W., Brinkmann J., Brownstein J., Eisenstein D., et al., \textit{Mon. Not. Roy. Astron. Soc.} \textbf{424} (2012) 2339.
\bibitem{Reid2012}
Reid B. A., Samushia L., White M., Percival W. J., Manera M., et al., \textit{Mon. Not. Roy. Astron. Soc.} \textbf{426} (2012) 2719.
\bibitem{delaTorre2013}
de la Torre S., Guzzo L., Peacock J., Branchini E., Iovino A., et al., \textit{Astron. Astrophys.} \textbf{557} (2013) A54.
\bibitem{Planck2015}
Planck 2015 results, XIII. Cosmological parameters, \textit{Astron. Astrophys.} \textbf{594} (2016) A13.
\bibitem{Hinshaw2013}  
WMAP collaboration, G. Hinshaw et al., \textit{Astrophys. J. Suppl.} \textbf{208} (2013) 19.
\bibitem{Beutler2011}
F. Beutler et al., \textit{Mon. Not. Roy. Astron. Soc.} \textbf{416} (2011) 3017.
\bibitem{Ross2015}
A. J. Ross et al., \textit{Mon. Not. Roy. Astron. Soc.} \textbf{449} (2015) 835.
\bibitem{Percival2010}
W. J. Percival et al., \textit{Mon. Not. Roy. Astron. Soc.} \textbf{401} (2010) 2148.
\bibitem{Kazin2010}
E. A. Kazin et al., \textit{Astrophys. J.} \textbf{710} (2010) 1444.
\bibitem{Padmanabhan2012}
N. Padmanabhan et al., \textit{Mon. Not. Roy. Astron. Soc.} \textbf{427} (2012) 2132.
\bibitem{Chuang2013a}
C. H. Chuang and Y. Wang, \textit{Mon. Not. Roy. Astron. Soc.} \textbf{435} (2013) 255.
\bibitem{Chuang2013b}
C-H. Chuang and Y. Wang, \textit{Mon. Not. Roy. Astron. Soc.} \textbf{433} (2013) 3559.
\bibitem{Anderson2014a}
L. Anderson et al., \textit{Mon. Not. Roy. Astron. Soc.} \textbf{441} (2014) 24.
\bibitem{Kazin2014}
E. A. Kazin et al., \textit{Mon. Not. Roy. Astron. Soc.} \textbf{441} (2014) 3524.
\bibitem{Debulac2015}  
T. Delubac et al., \textit{Astron. Astrophys.} \textbf{574} (2015) A59.
\bibitem{FontRibera2014} 
A. Font-Ribera et al., \textit{J. Cosmol. Astropart. Phys.} \textbf{05} (2014) 027.
\bibitem{Eisenstein1998}
D. J. Eisenstein, W. Hu, \textit{Astrophys. J.} \textbf{496} (1998) 605.
\bibitem{Eisenstein2005}
D. J. Eisenstein et al., \textit{Astrophys. J.} \textbf{633} (2005) 560.
\bibitem{Hemantha2014}
M. D. P. Hemantha, Y. Wang and C-H. Chuang., \textit{Mon. Not. Roy. Astron. Soc.} \textbf{445} (2014) 3737.
\bibitem{Bond-Tegmark1997}
J. R. Bond, G. Efstathiou and M. Tegmark, \textit{Mon. Not. Roy. Astron. Soc.} \textbf{291} (1997) L33.
\bibitem{Hu-Sugiyama1996}
W. Hu and N. Sugiyama, \textit{Astrophys. J.} \textbf{471} (1996) 542.
\bibitem{Neveu2016}
J. Neveu, V. Ruhlmann-Kleider, P. Astier, M. Besan\c con, J. Guy, A. M\"oller, E. Babichev, \textit{Astron. and Astrophys.} \textbf{600} (2017) A40.
\bibitem{Sharov2015}
G. S. Sharov, \textit{J. Cosmol. Astropart. Phys.} \textbf{06} (2016) 023.
\bibitem{Zhang2014} 
C. Zhang et al., \textit{Res. Astron. Astrophys.} \textbf{14} (2014) 1221.
\bibitem{Simon2005}
J. Simon, L. Verde and R. Jimenez, \textit{Phys. Rev.} \textbf{D 71} (2005) 123001.
\bibitem{Moresco2012}
M. Moresco et al., \textit{J. Cosmol. Astropart. Phys.} \textbf{8} (2012) 006.
\bibitem{Gastanaga2009}
E. Gasta\~naga, A. Cabre, L. Hui, \textit{Mon. Not. Roy. Astron. Soc.} \textbf{399} (2009) 1663.
\bibitem{Oka2014}
A. Oka et al., \textit{Mon. Not. Roy. Astron. Soc.} \textbf{439} (2014) 2515.
\bibitem{Blake2012}
C. Blake et al., \textit{Mon. Not. Roy. Astron. Soc.} \textbf{425} (2012) 405.
\bibitem{Stern2010}
D. Stern, R. Jimenez, L. Verde, M. Kamionkowski and S. A. Stanford, \textit{J. Cosmol. Astropart. Phys.} \textbf{02} (2010) 008.
\bibitem{Moresco2015}
M. Moresco, \textit{Mon. Not. Roy. Astron. Soc.} \textbf{450} (2015) L16-L20.
\bibitem{Busca2013}
N. G. Busca et al., \textit{Astron. Astrophys.} \textbf{552} (2013) A96.
\bibitem{DES2006}
V. Sahni, \textit{Lect. Notes Phys.} \textbf{653} (2004) 141; E. J. Copeland, M. Sami and S. Tsujikawa, \textit{Int. J. Mod. Phys.} \textbf{D 15} (2006) 1753.
\bibitem{OptionsDE}
S. Weinberg, \textit{Rev. Mod. Phys.} \textbf{61} (1989) 1;
V. Sahni and A. A. Starobinsky, \textit{Int. J. Mod. Phys.} \textbf{D 9} (2000) 373;
U. Seljak, et al., \textit{Phys. Rev.} \textbf{D 71} (2005) 103515;
E. Rozo et al., \textit{Astrophys. J.} \textbf{708} (2010) 645;
X. Cheng, Y. Gong and E. N. Saridakis, \textit{J. Cosmol. Astropart. Phys.} \textbf{04} (2009) 001;
Z. K. Guo, Y. S. Piao, X. M. Zang and Y. Z. Zhang, \textit{Phys. Lett.} \textbf{B 608} (2005) 177;
R. R. Caldwell, R. Dave, and P. J. Steinhardt, \textit{Phys. Rev. Lett.} \textbf{80} (1998) 1582;
C. Armendariz-Picon, V. Mukhanov, P. J. Steinhardt, \textit{Phys. Rev.} \textbf{D 63} (2001) 103510;
T. Chiba, T. Okabe, M. Yamaguchi, \textit{Phys. Rev.} \textbf{D 62} (2000) 023511;
M. K. Mak and T. Harko, \textit{Phys. Rev.} \textbf{D 71} (2005) 104022;
M. R. Garousi, M. Sami, and S. Tsujikawa, \textit{Phys. Rev.} \textbf{D 71} (2005) 083005.
\bibitem{Interacting}
Z. K. Guo, N. Ohta, and S. Tsujikawa, \textit{Phys. Rev.} \textbf{D 76} (2007) 023508; 
J. H. He and B. Wang, \textit{J. Cosmol. Astropart. Phys.} \textbf{06} (2008) 010;
S. Campo, R. Herrera and D. Pav\'on, \textit{J. Cosmol. Astropart. Phys.} \textbf{01} (2009) 020;
S. Cao, N. Liang and Z. H. Zhu, \textit{Int. J. Mod. Phys.} \textbf{D 22} (2013) 1350082;
Y. H. Li and X. Zhang, \textit{Eur. Phys. J. C.} \textbf{71} (2011) 1700;
\bibitem{Pavons}
D. Pav\'on, B. Wang, \textit{Gen.Rel.Grav.} \textbf{41} (2009) 1-5;
S. del Campo, R. Herrera, G. Olivares, and D. Pav\'on, \textit{Phys. Rev.} \textbf{D 74} (2006) 023501.
\bibitem{Wangs}
B. Wang, C. Y. Lin and E. Abdalla, \textit{Phys. Lett.} \textbf{B 637} (2006) 357;
R. G. Cai and Q. Su, \textit{Phys. Rev.} \textbf{D 81} (2010) 103514.
\bibitem{Cueva-Nucamendi2012}
F. Cueva Solano and U. Nucamendi, \textit{J. Cosmol. Astropart. Phys.} \textbf{04} (2012) 011;
F. Cueva Solano and U. Nucamendi, \textit{arXiv: 1207.0250} \textbf{07} (2012) 02.
\bibitem{valiviita2008}
J. Valiviita, E. Majerotto and R. Maartens, \textit{J. Cosmol. Astropart. Phys.} \textbf{07} (2008) 020.
\bibitem{Jackson2009}
B. M. Jackson, A. Taylor, and A. Berera, \textit{Phys. Rev.} \textbf{D 79} (2009) 043526.
\bibitem{He2009}
J. H. He, B. Wang, E. Abdalla, \textit{Phys. Lett.} \textbf{B 671} (2009) 139.
\bibitem{Xu2011}
X. D. Xu, J. H. He, B. Wang, \textit{Phys. Lett.} \textbf{B 701} (2011) 513.
\bibitem{He2011}
J. H. He, B. Wang, and E. Abdalla \textit{Phys. Rev.} \textbf{D 83} (2011) 063515.
\bibitem{Clemson2012}
T. Clemson, K. Koyama, G. B. Zhao, R. Maartens and J. Valiviita \textit{Phys. Rev.} \textbf{D 85} (2012) 043007.
\bibitem{Alcaniz2013}
S. Tsujikawa, A. De Felice and J. Alcaniz, \textit{J. Cosmol. Astropart. Phys.} \textbf{01} (2013) 030.
\bibitem{Yang2014}
W. Yang and L. Xu, \textit{Phys. Rev.} \textbf{D 89} (2014) 083517.
\bibitem{Chevallier-Linder}
M. Chevallier, D. Polarski, \textit{Int. J. Mod. Phys.} \textbf{D 10} (2001) 213;
E. V. Linder, \textit{Phys. Rev. Lett.} \textbf{90} (2003) 091301.  
\bibitem{Li-Ma}
H. Li and X. Zhang, \textit{Phys. Lett.} \textbf{B 703} (2011) 119;
J. Z. Ma and X. Zhang, \textit{Phys. Lett.} \textbf{B 699} (2011) 233. 
\bibitem{Martinez2008}
E. F. Martinez and L. Verde, \textit{J. Cosmol. Astropart. Phys.} \textbf{08} (2008) 023.
\bibitem{Campo-Herrera2015}
S. del Campo, R. Herrera, and D. Pav\'on \textit{Phys. Rev.} \textbf{D 91} (2015) 123539;
\bibitem{AIC_Criterion}
H. Akaike, \textit{IEEE Transactions on Automatic Control}, \textbf{19} (2003) 716; 
A. R. Liddle, \textit{Mon. Not. Roy. Astron. Soc.} \textbf{351} (2004) L49; 
M. Biesiada, \textit{J. Cosmol. Astropart. Phys.} \textbf{02} (2007) 003.
\bibitem{BIC_Criterion}
G. Schwarz, \textit{Annals of Statistics}, \textbf{6} (1978) 461; 
A. R. Liddle, \textit{Mon. Not. Roy. Astron. Soc.} \textbf{377} (2007) L74.
\bibitem{Kurek2014}
A. Kurek, M. Szydlowski, A. Krawiec, M.Kamionka, \textit{Eur. Phys. J.} \textbf{1-12} (2014). 
\bibitem{Arevalo2016}
F. Arevalo, A. Cid and J. Moya, \textit{Eur. Phys. J. C.} \textbf{77} (2017) 565. 
\bibitem{Koyama2009-Brax2010}
Kazuya Koyama, Roy Maartens, and Yong-Seon Song, \textit{J. Cosmol. Astropart. Phys.} \textbf{10} (2009) 017;
P. Brax, C. van de Bruck, D. F. Mota, N. J. Nunes, and H. A. Winther, \textit{Phys. Rev.} \textbf{D 82} (2010) 083503.
\bibitem{Ratios}
L. P. Chimento, A. S. Jakubi, D. Pav\'on, and W. Zimdahl, \textit{Phys. Rev.} \textbf{D 67} (2003) 083513;
J. Q. Xia and M. Viel, \textit{J. Cosmol. Astropart. Phys.} \textbf{04} (2009) 002.
\bibitem{Mas}
C. P. Ma and M. Sasaki, \textit{Prog. Theor. Phys. Suppl.} \textbf{78} (1984) 1;
Ma C. P. and Bertschinger E, \textit{Astrophys. J.} \textbf{455} (1995) 7.
\bibitem{Nesseris2007}
S. Nesseris and L. Perivolaropoulos, \textit{J. Cosmol. Astropart. Phys.} \textbf{01} (2007) 018.
\bibitem{Lu2009-Neveu2013}
S. Nesseris, A. De Felice, and S. Tsujikawa, \textit{Phys. Rev.} \textbf{D 82}, (2010) 124054;
J. Lu, \textit{Phys. Lett.} \textbf{B 680}, (2009) 404;
J. Neveu, et al., \textit{Astron. and Astrophys.} \textbf{555} (2013) A53.
\end{scriptsize}
  \end{thebibliography}
\end{document}